\documentclass{emulateapj}
\usepackage{graphicx}
\usepackage{epstopdf}
\newcommand{\asec}      {\mbox{$^{\prime \prime}  $} }
\newcommand{\amin}      {\mbox{$^{\prime}$}}
\begin{document}
\title{The First Release COSMOS Optical and Near-IR Data and Catalog$^{\star}$}
 \author{P. Capak\altaffilmark{1},
 H. Aussel\altaffilmark{2,47},  
M. Ajiki\altaffilmark{26},
 H. J. McCracken\altaffilmark{2,17},
B. Mobasher\altaffilmark{5},
N. Scoville\altaffilmark{1,2}, 
P. Shopbell\altaffilmark{1}, 
Y. Taniguchi\altaffilmark{26,45},
D. Thompson\altaffilmark{1,46},
S. Tribiano\altaffilmark{16,36},
S. Sasaki\altaffilmark{1,26,45},
A. W. Blain\altaffilmark{1},
M. Brusa\altaffilmark{13},
C. Carilli          \altaffilmark{6},
A. Comastri\altaffilmark{35},
C. M. Carollo\altaffilmark{8},
P. Cassata\altaffilmark{12},
J. Colbert\altaffilmark{31},
R. S. Ellis\altaffilmark{1},
M. Elvis\altaffilmark{10},
M. Giavalisco\altaffilmark{5},
W. Green\altaffilmark{1},
L. Guzzo\altaffilmark{12},
G. Hasinger\altaffilmark{13},
O. Ilbert\altaffilmark{4},
C. Impey\altaffilmark{14},
K. Jahnke\altaffilmark{25},
J. Kartaltepe\altaffilmark{4},
J-P. Kneib\altaffilmark{15},
J. Koda\altaffilmark{1},
A. Koekemoer\altaffilmark{5},
Y. Komiyama\altaffilmark{43},
A. Leauthaud\altaffilmark{1,15},
O. Lefevre\altaffilmark{15},
S. Lilly\altaffilmark{8},
R. Massey\altaffilmark{1},
S. Miyazaki\altaffilmark{18},
T. Murayama\altaffilmark{26},
T. Nagao            \altaffilmark{43,44},
J. A. Peacock\altaffilmark{32},
A. Pickles \altaffilmark{33},
C. Porciani\altaffilmark{8},
A. Renzini\altaffilmark{21,34},
J. Rhodes\altaffilmark{1,22},
M. Rich\altaffilmark{23},
M. Salvato\altaffilmark{1},
D. B. Sanders\altaffilmark{4},
C. Scarlata\altaffilmark{8},
D. Schiminovich\altaffilmark{24},
E. Schinnerer\altaffilmark{25},
M. Scodeggio\altaffilmark{38},
K. Sheth\altaffilmark{1,31},
Y. Shioya           \altaffilmark{45},
L. A. M. Tasca\altaffilmark{15},
J. E. Taylor\altaffilmark{1},
L. Yan\altaffilmark{31},
G. Zamorani\altaffilmark{29}}
 
\begin{abstract}
	We present imaging data and photometry for the COSMOS survey in 15 photometric bands between $0.3\mu m$ and $2.4\mu m$.  These include data taken on the Subaru 8.3m telescope, the KPNO and CTIO 4m telescopes, and the CFHT 3.6m telescope. Special techniques are used to ensure that the relative photometric calibration is better than 1\% across the field of view.  The absolute photometric accuracy from standard star measurements is found to be 6\%.  The absolute calibration is corrected using galaxy spectra, providing colors accurate to 2\% or better.  Stellar and galaxy colors and counts agree well with the expected values.  Finally, as the first step in the scientific analysis of these data we construct panchromatic number counts which confirm that both the geometry of the universe and the galaxy population are evolving.

\end{abstract}


\altaffiltext{$\star$}{Based in part on observations with :   The NASA/ESA Hubble Space Telescope, obtained at the Space Telescope Science
Institute, which is operated by AURA Inc, under NASA contract NAS5-26555.  The Subaru Telescope, which is operated by the National Astronomical Observatory of Japan.  The MegaPrime/MegaCam, a joint project of CFHT and CEA/DAPNIA, at the Canada-France-Hawaii Telescope (CFHT) which is operated by the National Research Council (NRC) of Canada, the Institute National des Science de l'Univers of the Centre National de la Recherche and  the University of Hawaii.  The Kitt Peak National Observatory, Cerro Tololo Inter-American Observatory and the National Optical Astronomy Observatory, which is operated by the Association of Universities for Research in Astronomy Inc. (AURA) under cooperative agreement with the National Science Foundation.}  

\altaffiltext{1}{California Institute of Technology, MC 105-24, 1200 East California Boulevard, Pasadena, CA 91125}
\altaffiltext{2}{Visiting Astronomer, Univ. Hawaii, 2680 Woodlawn Dr., Honolulu, HI, 96822}
\altaffiltext{3}{Canadian Institute for Theoretical Astrophysics, Mclennan Labs, University of Toronto, 60 St. George St, Room 1403, Toronto, ON M5S 3H8, Canada}
\altaffiltext{4}{Institute for Astronomy, 2680 Woodlawn Dr., University of Hawaii, Honolulu, Hawaii, 96822}
\altaffiltext{5}{Space Telescope Science Institute, 3700 San Martin Drive, Baltimore, MD 21218}
\altaffiltext{6}{National Radio Astronomy Observatory, P.O. Box 0, Socorro, NM 87801-0387}
\altaffiltext{7}{Department of Physics, University of Chicago, 5640 South Ellis Avenue, Chicago, IL 60637}
\altaffiltext{8}{Department of Physics, ETH Zurich, CH-8093 Zurich, Switzerland}
\altaffiltext{9}{National Optical Astronomy Observatory, P.O. Box 26732, Tucson, AZ 85726}
\altaffiltext{10}{Harvard-Smithsonian Center for Astrophysics, 60 Garden Street, Cambridge, MA 02138}
\altaffiltext{11}{Department of Physics, Carnegie Mellon University, 5000 Forbes Avenue, Pittsburgh, PA 15213}
\altaffiltext{12}{INAF-Osservatorio Astronomico di Brera, via Bianchi 46, I-23807 Merate (LC), Italy}
\altaffiltext{13}{Max Planck Institut f\"{u}r Extraterrestrische Physik,  D-85478 Garching, Germany}
\altaffiltext{14}{Steward Observatory, University of Arizona, 933 North Cherry Avenue, Tucson, AZ 85721}
\altaffiltext{15}{Laboratoire d'Astrophysique de Marseille, BP 8, Traverse du Siphon, 13376 Marseille Cedex 12, France}
\altaffiltext{16}{American Museum of Natural History, Central Park West at 79th Street, New York, NY  10024}
\altaffiltext{17}{Institut d'Astrophysique de Paris, UMR7095 CNRS, Universit\`{e} Pierre et Marie Curie, 98 bis Boulevard Arago, 75014 Paris, France}
\altaffiltext{18}{Subaru Telescope, National Astronomical Observatory of Japan, 650 North Aohoku Place, Hilo, HI 96720}
\altaffiltext{19}{Department of Physics and Astronomy, Johns Hopkins University, Homewood Campus, Baltimore, MD 21218}
\altaffiltext{20}{Service d'Astrophysique, CEA/Saclay, 91191 Gif-sur-Yvette, France}
\altaffiltext{21}{European Southern Observatory, Karl-Schwarzschild-Str. 2, D-85748 Garching, Germany}
\altaffiltext{22}{Jet Propulsion Laboratory, Pasadena, CA 91109}
\altaffiltext{23}{Department of Physics and Astronomy, University ofCalifornia, Los Angeles, CA 90095}
\altaffiltext{24}{Department of Astronomy, Columbia University, MC2457, 550 W. 120 St. New York, NY 10027}
\altaffiltext{25}{Max Planck Institut f\"{u}r Astronomie, K\"{o}nigstuhl 17, Heidelberg, D-69117, Germany}
\altaffiltext{26}{Astronomical Institute, Graduate School of Science, Tohoku University, Aramaki, Aoba, Sendai 980-8578, Japan}
\altaffiltext{27}{Department of Astronomy, Yale University, P.O. Box 208101, New Haven, CT 06520-8101}
\altaffiltext{29}{INAF-Osservatorio Astronomico di Bologna, via Ranzani 1, I-40127 Bologna, Italy}
\altaffiltext{30}{Max-Planck-Institut f\"{u}r Astrophysik, D-85748 Garching bei M\"{u}nchen, Germany}
\altaffiltext{31}{Spitzer Science Center, California Institute of Technology, Pasadena, CA 91125}
\altaffiltext{32}{Institute for Astronomy, University of Edinburgh, Royal Observatory, Blackford Hill, Edinburgh EH9 3HJ, U.K.}
\altaffiltext{33}{Caltech Optical Observatories, MS 320-47, California Institute of Technology, Pasadena, CA 91125}
\altaffiltext{34}{Dipartimento di Astronomia, Universitˆ di Padova, vicolo dell'Osservatorio 2, I-35122 Padua, Italy}
\altaffiltext{35}{INAF-Osservatorio Astronomico di Bologna, via Ranzani 1, 40127 Bologna, Italy}
\altaffiltext{36}{CUNY Borough of Manhattan Community College, 199 Chambers St., New York, NY 10007}
\altaffiltext{37}{Astrophysical Observatory, City University of New York, College of Staten Island, 2800 Victory Blvd, Staten Island, NY  10314}
\altaffiltext{38}{INAF - IASF Milano, via Bassini 15, 20133 Milano, Italy}
\altaffiltext{39}{Laboratoire d'Astrophysique de Toulouse et de Tarbes (LA2T - UMR 5572), Observatoire Midi-PyrŽnŽes, 14 avenue E. Belin, F-31400 Toulouse, France}
\altaffiltext{40}{Observatoire de Paris, LERMA, 61 Avenue de l'Observatoire, 75014 Paris, France}
\altaffiltext{41}{Physics Department, Graduate School of Science, Ehime University, 2-5 Bunkyou, Matuyama, 790-8577, Japan}
\altaffiltext{42}{Large Binocular Telescope Observatory, University of Arizona, 933 N. Cherry Ave., Tucson, AZ  85721-0065,   USA}
\altaffiltext{43}{National Astronomical Observatory of Japan, 2-21-1 Osawa, Mitaka, Tokyo 181-8588, Japan}
\altaffiltext{44}{INAF -- Osservatorio Astrofisico di Arcetri, Largo Enrico Fermi 5, 50125 Firenze, Italy}
\altaffiltext{45}{Physics Department, Graduate School of Science \& Engineering, Ehime University, 2-5 Bunkyo-cho, Matsuyama, 790-8577, Japan}       
\altaffiltext{46}{LBT Observatory, University of Arizona, 933 N. Cherry Ave., Tucson, Arizona, 85721-0065, USA}
\altaffiltext{47}{AIM Ð UnitŽ Mixte de Recherche CEA Ð CNRS Ð UniversitŽ Paris VII Ð UMR n¡ 7158}

\section{Introduction}
	Advances in astronomy are often driven by improved accuracy and precision along with increases in sensitivity and area of the available data.  The Canada-France Redshift Survey (CFRS)\citep{1995ApJ...455...50L}, the Hawaii Deep Surveys (HDSs) \citep{1999AJ....118..603C}, and the Hubble Deep Fields (HDFs) \citep{1996AJ....112.1335W, 2000AJ....120.2747C} were the first deep imaging and spectroscopic surveys aimed at understanding galaxy formation and evolution.  These discovered the global decline in star formation at $z<1$ and showed that this was due to star formation occurring in smaller galaxies at later times \citep{1996ApJ...460L...1L,1999AJ....118..603C}, a phenomenon often referred to as "Cosmic Downsizing".  At the same time \citet{1996AJ....112..352S,1999ApJ...519....1S,2003ApJ...592..728S} used the Lyman-Break Galaxy (LBG) color selection technique to identify galaxies at high redshift, dramatically improving the efficiency of spectroscopic surveys at $z>3$. Other selections such as the BzK \citep{2004ApJ...617..746D}, BX/BM\citep{2004ApJ...607..226A}, and DRG (Distant Red Galaxy) \citep{2003ApJ...587L..79F} have allowed for efficient sorting of $1<z<3$ galaxies. 

	Photometric redshifts are the logical extension of color selection by estimating redshifts and spectral energy distributions (SEDs) from many photometric bands.  Unlike color selection, photometric redshifts take advantage of all available information, enabling redshift estimates along with the age, star formation rate (SFR) and mass.  Unfortunately, photometric redshifts are also susceptible to systematics in all bands.  This increases the calibration requirements, especially the required photometric accuracy, for modern cosmological surveys such as the Great Observatories Origins Deep Survey (GOODS) \citep{2004ApJ...600L..93G}, the Galaxy Evolution from Morphology and Spectral Energy Distributions (GEMS) survey \citep{2004ApJS..152..163R}, and the Cosmic Evolution Survey or COSMOS \citep{scoville_cosmos_overview}.
	
	 GOODS and GEMS are designed to study evolution of galaxies with look back time, whereas COSMOS is designed to probe the evolution of galaxies in the context of their large-scale structure out to moderate redshift.  The desire to study large-scale structure in COSMOS necessitates a 2 square degree area with deep pan-chromatic data.  Such data have been collected at nearly every observable wavelength from the X-rays to the radio.  The study of large-scale structures places strong calibration requirements on the COSMOS data;  for example spatial variations in photometry and astrometry must be kept to a minimum, typically less than 1\% for photometry to ensure high quality photometric redshifts and 0.01\asec for astrometry to enable measurements of weak lensing and correlation functions.  Meeting these calibration requirements is often difficult as multiple instrument pointings are used to cover the field.
	 
	 This paper concentrates on the ground-based data reduction, the multi-band optical and near-infrared catalog and the steps taken to ensure a high level of photometric consistency.  The observing strategy for the Subaru Suprime-Cam observations, which form the bulk of our ground based data, are discussed separately in \citet{tanaguchi-cosmos}.  In addition, the absolute photometric and astrometric system used here is defined in \citet{aussel-photom}.  

	 An overview of the COSMOS project and its goals are given in \citet{scoville_cosmos_overview}.  Details of the Hubble Space Telescope (HST) observations, including the Advanced Camera for Surveys (ACS), the Wide Field Planetary Camera 2 (WFPC2), and the Near Infrared Camera and Multi-Object Spectrometer (NICMOS) are found in \citet{scoville-hst}.  The ACS data acquisition and reduction are detailed in \citet{koekemore-hst}, and a monochromatic catalog based only on the HST-ACS observations is presented in \citet{acs-weak-lensing}.  Observations at other wavelengths consist of: X-ray observations with XMM \citep{hasinger_xmmcosmos}, ultraviolet (UV) observations with GALEX \citep{zamojski_cosmos_galex}, mid-infrared observations with the Spitzer Space Telescope \citep{sanders_cosmos_spitzer}, sub-mm observations from the Caltech Sub-mm Observatory (CSO) \citep{aguirre07} and Institut de Radioastronomie Millim\'{e}trique (IRAM) 30m telescope \citep{bertoldi_cosbo}, and radio observations with the Very Large Array (VLA) \citep{2004AJ....128.1974S,schinnerer_cosmos_vla2}.
	 
	 We begin by presenting an overview of the various data sets and photometric systems, the imaging data products and the data reduction in Section \ref{s:data-redux}.  PSF matching is covered in Section \ref{s:psf-match}, and the generation of a multi-color catalog is presented in Section \ref{s:catalog}.  Finally, in Section \ref{s:discussion} we conduct several quality checks, and suggest several corrections to the absolute photometry.
	 
\section{Observations and Data Reduction}\label{s:data-redux}

\begin{figure}
\begin{center}
\includegraphics[scale=0.8]{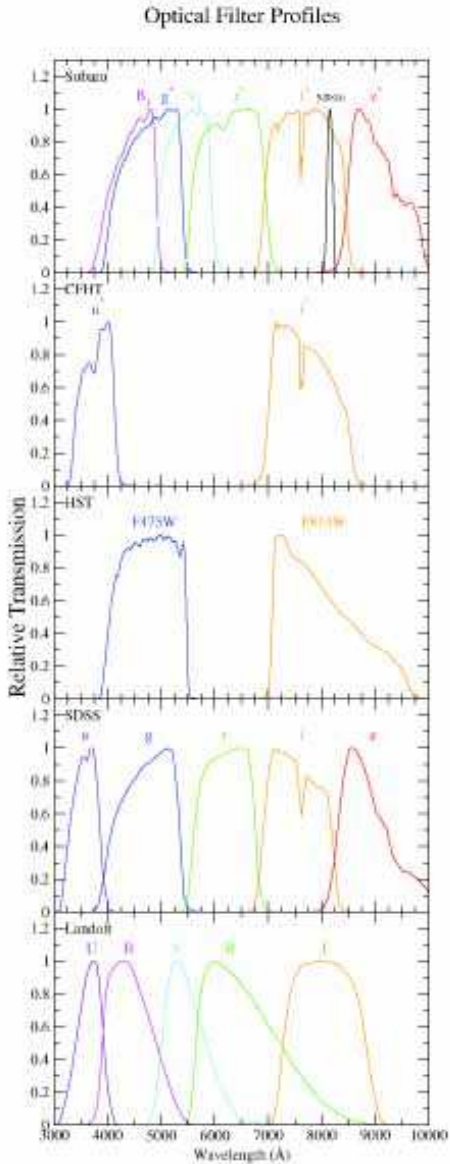}
\caption{Filter transmission profiles are shown for the COSMOS optical data set from CFHT, Subaru, and HST as of April, 2005.  These profiles are normalized to a maximum throughput of one and include the transmission of the Atmosphere, the telescope, the camera optics, the filter, and the detector.  The HST F475W data only covers the central $9\amin \times 9\amin$, details are given in \citet{scoville-hst}. The SDSS \citet{SDSS-DR2} and Johnson-Cousins system used by \citet{Landolt} are shown for comparison.  Notice the significant differences between the Johnson-Cousins, SDSS, and other systems.  Color conversions are clearly needed to transform from the COSMOS system to the standard star systems.  These are given in \citet{aussel-photom}. \label{f:opt-filter}}
\end{center}
\end{figure}

\begin{figure}
\begin{center}
\includegraphics[scale=0.3]{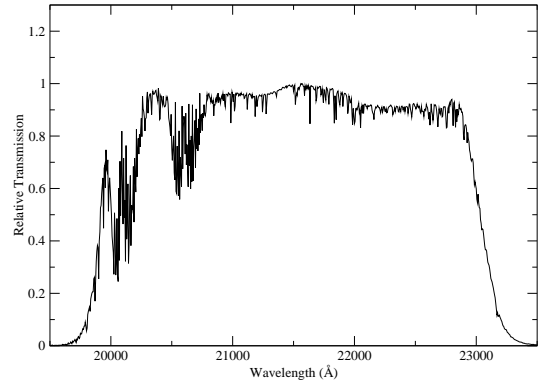}
\caption{The transmission profile of the $K_s$ band filter from KPNO-FLAMINGOS and CTIO-ISPI is shown. This profile is normalized to a maximum throughput of one and includes the transmission of the atmosphere, the telescope, the camera optics, the filter, and the detector.\label{f:K-filter}}
\end{center}
\end{figure}

	The present COSMOS data were collected on a variety of telescopes and instruments, as well as from the Sloan Digital Sky Survey (SDSS) second data release (DR2) archive  \citep{SDSS-DR2}.  This paper covers the processing of the data obtained with Suprime-Cam \citep{suprimecam} on the Subaru 8.3m telescope, Megaprime \citep{megaprime,2003SPIE.4841...72B} on the Canada France Hawaii 3.6m Telescope (CFHT), Flamingos \citep{flamingos} on the Kitt Peak National Observatory 4m (KPNO-4m), and the Infrared Side Port Imager (ISPI) \citep{ISPI} on the Cerro Tololo International Observatory 4m (CTIO-4m) during the 2004--2005 observing season.   

\begin{figure}
\begin{center}
\includegraphics[scale=0.3]{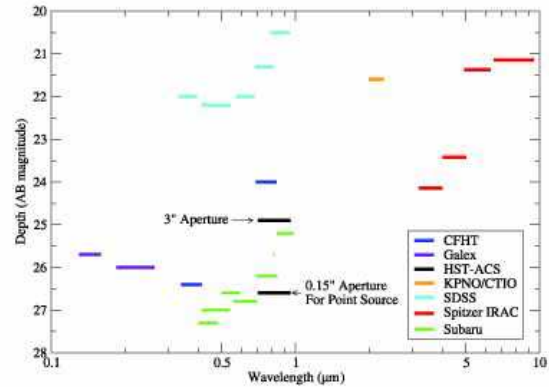}
\caption{The background limited depth of COSMOS observations in the Ultraviolet, Optical, and Infrared are shown. CFHT, KPNO/CTIO, and Subaru depths are $5\sigma$ in a 3\asec aperture.  SDSS depths are those quoted in \citet{SDSS-DR2}. The depth of the HST-ACS observations is given for a 3\asec aperture and a 0.15\asec aperture with a point source.  A 3\asec aperture is optimal for color measurements, while the 0.15\asec aperture is the $5\sigma$ detection limit for point sources.  The Spitzer-IRAC depths are those expected at $5\sigma$ in a 3\asec aperture for observations taken in 2006.  The GALEX depths are from \citet{zamojski_cosmos_galex}. \label{f:depth}}
\end{center}
\end{figure}

	The telescopes and instruments used for the COSMOS survey are presented in Table \ref{t:telescopes}.   A survey efficiency is given for each telescope-instrument pair to allow comparisons between the various data sets.  The survey efficiency is defined as the telescope collecting area multiplied by the detector imaging area (degrees$^2$ m$^2$) and does not include variations in detector sensitivity, sky background, or field geometry.  This number is most useful for comparing observations taken in similar bands.  The filter transmission profiles, including atmospheric transmission, telescope reflectivity, instrument optical transmission, filter transmission and detector sensitivity are plotted in Figures \ref{f:opt-filter} \& \ref{f:K-filter} in units of relative detector quantum efficiency normalized to 1 at the peak.

\begin{deluxetable*}{lllcccl}
\tabletypesize{\scriptsize}
\tablecaption{Telescopes used for COSMOS optical/IR data in 2005-2006\label{t:telescopes}}
\tablehead{
\colhead{Telescope}&\colhead{Telescope}&\colhead{Instrument}&\colhead{Field of }& \colhead{Instrument Wavelength}&\colhead{Survey \tablenotemark{1}} & \colhead{Filters}\\
\colhead{}		       &\colhead{Diameter} &\colhead{}		 &\colhead{View}     & \colhead{Sensitivity}                     &\colhead{Efficiency}                              & \colhead{used}}
\startdata
CFHT	&	3.6m		&	Mega-Prime	&	$56.4\amin \times 57.6\amin$  & 3200-11000\AA &   9.19    &   $u^*$, $i^*$\\
CTIO	&	4m		&	ISPI			&	$10.2\amin \times 10.2\amin$  & 0.9-2.5$\mu m$ & 0.37    &    $K_s$\\
HST		&	2.5m		&	ACS-WFC	&	$3.4\amin \times 3.4\amin$       & 4000-11000\AA &  0.02    &   $F814W$\\
KPNO	&	4m		&	FLAMINGOS	&	$10.8\amin \times 10.8\amin$  & 0.9-2.5$\mu m$  & 0.41    &   $K_s$\\
SDSS	&	2.5m		&	SDSS		&	$25 \times 13.5\amin \times 13.5 \amin$ & 3200-11000\AA & 2.49 & $u$, $g$, $r$, $i$, $z$\\
Subaru	&	8.3m		&	Suprime-Cam	&	$34\amin \times 27 \amin$        & 4000-11000\AA & 13.8     &   $B_J$, $V_J$, $g^+$, $r^+$,\\
         	&			&				&						        & 				&      &   $i^+$, $z^+$, $NB816$\\
\enddata
\tablenotetext{1}{Defined as the telescope collecting area multiplied by the imaging area in square degrees.}
\end{deluxetable*}
	
	The Suprime-Cam, Megaprime, SDSS Photometric and SDSS Survey cameras have filters distinct from each other and from the Landolt standard star system.  Even the SDSS photometric telescope and SDSS Survey telescope filter sets differ from one another by 2-4\%.  To differentiate between these filter systems we use a $^+$ superscript for the Suprime-Cam Sloan filters and a $^*$ superscript for the Megaprime Sloan filters; no superscript is used for the SDSS survey filters.  The designation $U$, $B$, $V$, $R$, and $I$ are used for the Landolt-Johnson-Cousins set while $B_J$ and $V_J$ are used for the Suprime-Cam Johnson set. Conversions between these systems are discussed in \citet{aussel-photom}. 
			
	The wavelength range, depth, and image quality for all data presently reduced and included in the version 2.0 optical/IR catalog are given in Table \ref{t:data-depth} and plotted in Figure \ref{f:depth}.  The depth quoted is for a $5\sigma$ measurement in a 3\asec aperture of an isolated point source at the median seeing given in Table \ref{t:data-depth}.  This should be viewed as an optimistic estimate since most objects are extended and many are confused with neighboring sources.  \citet{tanaguchi-cosmos} present a discussion of detection sensitivities and completeness for various Subaru filters. The median photometric depths in the COSMOS $i^+$ selected catalog are discussed in Section \ref{s:catalog}.  Table \ref{t:data-depth} also gives a first order offset to the Vega system; however, a color term must be applied to get the true Landolt-Vega system magnitudes--these are given in \citet{aussel-photom}.  
	
\subsection{Data Products }\label{s:data-products} 

\begin{figure}
\begin{center}
\includegraphics[scale=0.3]{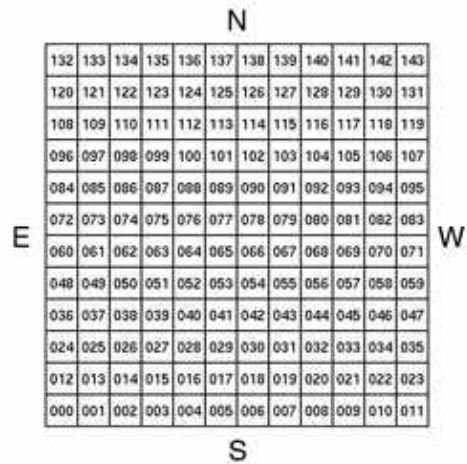}
\caption{The layout of image tiles for the COSMOS field.\label{f:tile-layout}}
\end{center}
\end{figure}

	We took special care in producing data products that simplify analysis and are tractable on contemporary computers\footnote{All data products discussed in this paper are publicly available at \url[http://irsa.ipac.caltech.edu/data/COSMOS/]{http://irsa.ipac.caltech.edu/data/COSMOS/}}.  To do this we defined a common grid of sub-images for all data products.  The starting point for this grid is the COSMOS astrometric catalog that covers 4 sq degrees \citep{aussel-photom} and is larger than all present or planned COSMOS data sets.  The area is divided into 144 sections of $10\amin \times 10\amin$, each section is covered by an image of size $4096 \times 4096$ pixels with a pixel scale of 0.15\asec.  Therefore, adjacent tiles overlap each other by 14.4\asec on all sides.  As a result, the vast majority of objects can be analyzed on a single image.  The layout of the image tiles is shown in Figure \ref{f:tile-layout}.  The pixel scale was chosen to be an integer multiple of the 0.05\asec scale used for the HST-ACS images.  All images and noise maps are scaled to units of nJy per pixel, which corresponds to a magnitude zero point of 31.4.  
	
	For each Subaru and SDSS band an image with the original Point Spread Function (PSF) and a PSF homogenized across the field within that band is provided.  For the Subaru $B_J$, $r^+$, and $i^+$ bands, which have exceptional image quality (0.5\asec- 0.8\asec seeing), a "best seeing" image is also provided.  The CFHT images were taken in queue observing mode, ensuring a consistent PSF for all observations, so only an original PSF image is provided for these data.  Finally, due to the large variation of the PSF in the CTIO and KPNO data, only a PSF homogenized image is provided.

	RMS noise maps are also provided for each filter.  These are on the same tiling scheme and flux scale as the images.  The RMS maps include noise contributions from photon noise, background subtraction, flat fielding, defect masking, saturation, and cosmic ray removal.  They do not include the photon noise contribution from object flux.
	
\begin{deluxetable*}{ccccccc}
\tabletypesize{\scriptsize}
\tablecaption{Data Quality and Depth\label{t:data-depth}}
\tablehead{
\colhead{Filter}          & \colhead{Central}                  & \colhead{Filter}            & \colhead{Seeing}             & \colhead{Depth\tablenotemark{1,2}}  & \colhead{Saturation\tablenotemark{2}} & \colhead{Offset from \tablenotemark{3}}\\
\colhead{Name}         & \colhead{Wavelength (\AA)} & \colhead{Width (\AA)} & \colhead{Range (\asec)}  & \colhead{}                                          & \colhead{Magnitude}                              & \colhead{Vega System}
}
\startdata
$u$			&	3591.3	&	550		&	1.2-2.0	& 22.0 & 12.0 & 0.921\\
$u^*$		&	3797.9	&	720		&	0.9		& 26.4 & 15.8 & 0.380\\
$B_J$		&	4459.7	&	897		&	0.4-0.9	& 27.3 & 18.7 & -0.131\\
$g$			&	4723.1	&	1300		&	1.2-1.7	& 22.2 & 12.0 & -0.117\\
$g^+$		&	4779.6	&	1265		&	0.7-2.1	& 27.0 & 18.2 & -0.117\\
$V_J$		&	5483.8	&	946		&	0.5-1.6	& 26.6 & 18.7 & -0.004\\
$r$			&	6213.0	&	1200		&	1.0-1.7	& 22.2 & 12.0 & 0.142\\
$r^+$		&	6295.1	&	1382		&	0.4-1.0	& 26.8 & 18.7 & 0.125\\
$i$			&	7522.5	&	1300		&	0.9-1.7	& 21.3 & 12.0 & 0.355\\
$i^+$		&	7640.8	&	1497		&	0.4-0.9	& 26.2 & 20.0\tablenotemark{*} & 0.379\\
$i^*$			&	7683.6	&	1380		&	0.94		& 24.0 & 16.0 & 0.380\\
$F814W$		&	8037.2	&	1862		&	0.12		& 24.9\tablenotemark{+} & 18.7 & 0.414\\
$NB816$		&	8151.0	&	117		&	0.4-1.7	& 25.7 & 16.9 & 0.458\\
$z$			&	8855.0	&	1000		&	1-1.7		& 20.5 & 12.0 & 0.538\\
$z^+$		&	9036.9	&	856		&	0.5-1.1	& 25.2 & 18.7 & 0.547\\
$K_s$		&	21537.2	&	3120		&	1.3		& 21.6 & 10.0 & 1.852\\
\enddata
\tablenotetext{1}{$5\sigma$ in a 3\asec aperture for an isolated point source at the native seeing.}
\tablenotetext{2}{In AB magnitudes.}
\tablenotetext{3}{AB magnitude = Vega Magnitude + Offset.  This offset does not include the color conversions to the Johnsons-Cousins system used by \citet{Landolt}.}
\tablenotetext{*}{Compact objects saturate at $i^+<21.8$ due to the exceptional seeing.}
\tablenotetext{+}{The sensitivity for photometry of an optimally extracted point source is 27.1,  for optimal photometry of a 1\asec galaxy it is  26.1}
\end{deluxetable*}

\subsection{Subaru Suprime-Cam }\label{s:subaru-redux}

	The Suprime-Cam instrument \citep{suprimecam} on the Subaru 8.3m telescope has a $34\amin \times 27\amin$ field of view. The camera has 10 2k $\times$ 4k Lincoln Labs CCD detectors which have good sensitivity between 4000\AA\ and 10000\AA.   Nine Suprime-Cam pointings were required to cover the COSMOS field.  During 2004 and 2005, data were obtained in the B$_J$, V$_J$, g$^+$, r$^+$, i$^+$, and z$^+$ broad-band and the NB816 narrow-band filters.  These Suprime-Cam observations, which required special planning, are detailed in \citet{tanaguchi-cosmos}.  Further observations in eleven 300\AA\ intermediate bands, IA427, IA464, IA484, IA505, IA527, IA624, IA679, IA709, IA738, IA767, IA827, and one narrow band NB711 were obtained in 2006 and 2007.  These new observations have been reduced using the prescription described here, but will be presented elsewhere.  
		
	 Objects brighter than 19$^{th}$ magnitude are saturated in a typical exposure and under good seeing the saturation level can drop to 22$^{nd}$ magnitude in long exposures.   As a result, it is extremely difficult to astrometrically calibrate these data against external astrometric catalogs such as the SDSS \citep{SDSS-DR2} and USNO-B1.0 \citep{usnob} which only reach 21$^{st}$ magnitude.  To mitigate this limitation, a series of short exposures were taken in each band. 

\subsubsection{Initial Calibration }\label{s:subaru-initcal}

	The first step in our Suprime-Cam data reduction is to measure a bias level from the over-scan region and subtract it from all the images.  Then all  bad or saturated pixels are masked.  Next, a median bias frame is constructed from over-scan corrected frames.  Following the over-scan correction, this bias frame is subtracted from data and flat frames in order to remove bias structures.  In particular, the bias level increases near the edges of the CCDs farthest from the readout register. 
	
	A median dome flat for each band is then constructed from 10-20 bias subtracted flat-field images.  The median dome flats and median biases are inspected for bad pixels, charge traps, and other defects that need to be masked. The appropriate median dome flat, with all defects masked, is then used to normalize all data frames.  Finally portions of the image vignetted by the guide probe are masked from all data frames using the position of the guide probe recorded in the image header.
	
	After the initial calibration catalogs are generated for every data frame with the IMCAT\footnote{\url[http://www.ifa.hawaii.edu/~kaiser/imcat/]{http://www.ifa.hawaii.edu/$\sim$kaiser/imcat/}} `hfindpeaks' and `apphot' routines using 5\asec diameter apertures.  The large aperture is chosen to minimize photometric variations caused by changes in seeing.   This catalog is then used to generate an object mask for each frame and to calculate the astrometric solution.
	
	The night sky is subtracted in a two--step process to account for fringing and scattered light.  First, a normalized median sky frame is constructed for each night.  The median frame is generated by masking objects in all frames, normalizing every frame to the same median flux, and finally median combining the normalized images.  The median sky is then scaled to the median background level in each frame and subtracted.  This removes both night sky illumination and fringing.  
	
	After subtracting the median sky residual scattered light is visible on the images.  This residual light affects both the overall flat-field and the background of each individual frame.  A correction to the flat-field is described in Section \ref{s:scat-light}, while the light in each frame is subtracted by masking objects and measuring the median of the residual background in  $128 \times 128$ pixel squares.  A background image is then generated by tesselating  over the grid of medians.   After subtracting this background, no visible sky structure is left on the individual frames.  However, this step creates negative halos around bright stars and very extended galaxies due to imperfect masking.  Fortunately, the amplitude of these haloes is similar in all frames and can be accounted for as a residual background in the combined images.  
	
	After sky subtraction, an astrometric solution is calculated separately for all exposures and CCDs by matching the object catalogs to the COSMOS astrometric catalog \citep{aussel-photom} using a 4th order two-dimensional polynomial.  The polynomial fits are improved by removing mismatched objects in an iterative fashion until the solution converges (typically in 2 iterations).  The resulting scatter between the fit positions and the final astrometry is always less than 0.2\asec at the $1\sigma$ level, independent of position.
		
	Using the astrometric solutions, defects around charge bleeds from saturated stars are masked using a list of bright stars from the SDSS and the USNO-B1.0.   Cosmic rays events are removed by detecting sharp edges in the images.  Finally, every frame is visually inspected to remove internal reflections, satellites, asteroids, and other false objects.  Once all masking is complete a new photometric catalog is generated containing only isolated objects in unmasked regions.

\subsubsection{Scattered Light Correction }\label{s:scat-light}

	Mechanical and optical constraints make it impossible to baffle wide field cameras against all scattered light.   The scattered light is equivalent to an unknown dark current added to each image, and must be subtracted rather than divided out.  As a result the usual flat fielding technique of observing a uniform light source such as the dome or sky is inaccurate at the 3-5\% level.  					

\begin{figure}
\begin{center}
\includegraphics[scale=0.3]{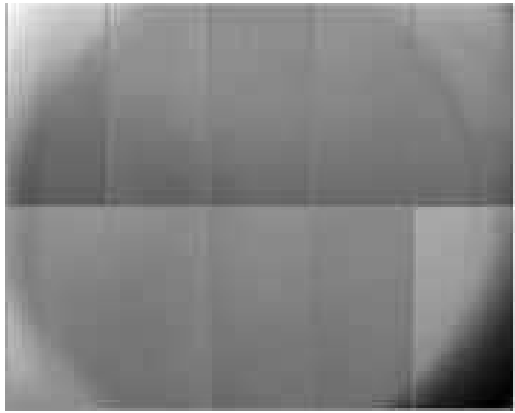}
\caption{The difference between Suprime-Cam dome flats taken in $r^+$ at position angles of 0 and +90 degrees is shown with chip to chip sensitivity variations removed.  The scale is linear with a stretch of -3\% to +3\% from black to white.  Variations in the illumination pattern due to scattered light are clearly visible. \label{f:dome-flat-diff}}
\end{center}
\end{figure}

	For Suprime-Cam the scattered light pattern and strength change significantly with the lighting conditions and telescope position. Variations as large as $\pm5$\% are observed at the edges of the field between dark, twilight, and dome conditions.  Figure \ref{f:dome-flat-diff} shows the difference between two dome flats taken at different rotation angles.  This effect is similar in amplitude and pattern to that observed with the 12k and Megacam cameras on the CFHT\footnote{\url[http://www.cfht.hawaii.edu/Instruments/Elixir/scattered.html]{http://www.cfht.hawaii.edu/Instruments/Elixir/scattered.html}}.
		
	Following the example of CFHT we calculate the true flat by observing objects at multiple positions on the camera.  The true flat can then be solved for as the flat-field, which yields the same background subtracted flux for an object at any position in the field of view. In practice the flat image is generated by dividing the focal plane into $128 \times 128$ pixel regions $r$, and calculating a factor $C_r$ for each region.  The regions are defined so that no region crosses a detector boundary.  As a result, sensitivity variations due to detectors are also measured.  We can also allow for an additional factor $P_e$ between exposures to correct for photometric variations due to atmospheric conditions and seeing.  However, if the data are photometric and corrected for airmass, $P_e=0$ for all exposures.
	
	For a single object the real magnitude $M_{real}$ is described by 
	
\begin{equation}
M_{real} = M + C_r + P_e
\label{e:single-object}
\end{equation}

{\noindent where $M$ is the measured magnitude.  If we consider a pair of exposures, a and b, we can construct a $\chi^2$ relation as: }

\begin{equation}
\chi^2 = \sum^{Nexp}_{a=0}\sum^{Nexp}_{b=a+1}\sum^{Nobj}_{i=o} \frac{(M_{i,a} - M_{i,b} + C_{r,a} - C_{r,b} + P_{a} - P_{b})^2}{\sigma_{i,a}^2 + \sigma_{i,b}^2}
\label{e:flat-gen}
\end{equation}
		
{\noindent which can be minimized to obtain the $C_r$ and $P_e$ factors.  Since an object can only belong to one region in each exposure we use the notation $C_{r,a}$ to indicate the region an object belongs to in exposure $a$.}
		
	To ensure that Equation \ref{e:flat-gen} is well constrained a series of 24 short exposures were taken in each band in each of two orientations.  Each exposure overlapped its neighbors by 50\%.  As a result, any given point in the COSMOS field was observed by at least eight different areas of Suprime-Cam.  These data were calibrated and the photometry performed as described in section \ref{s:subaru-initcal}.  The photometry was then corrected for variations in pixel area using the Jacobian of the optical distortion.  This corrects for the assumption that all pixels have the same size when the dome flat was applied.  Each object is then assigned a unique ID by merging the catalogs from each exposure with the master astrometry catalog.  The IMCAT routine `fitmagshifts' was then used to solve Equation \ref{e:flat-gen} for the $P_e$ and $C_r$ factors. 	
	
	Two Flat-fields were generated for each band, all the data was used in one, and only the data with high photometric quality in another.  No significant difference was observed between these flats in any band.  As a result we included all the data in our solutions.  This provides an additional check on the photometric calibration because an airmass term can be calculated independently from the data and from the standard stars.  In addition non-photometric data can be properly corrected for extinction.
		
\begin{figure}
\begin{center}
\includegraphics[scale=0.3]{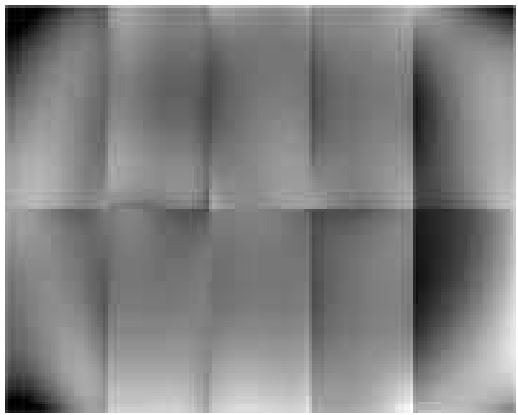}
\caption{The relative correction to the $r^+$ Suprime-Cam dome flat is shown with chip to chip sensitivity variations removed.  The scale is linear with a stretch of -3\% to +3\% from black to white.  A correction for scattered light in the vignetted portion of the field is clearly visible around the edge of the field of view. \label{f:dome-flat-corr}}
\end{center}
\end{figure}
\clearpage

	The 24 short exposures were taken while the camera was under similar but opposite gravity loads for each orientation.  The flat-fields measured from these two orientations agree to within 1\%, ruling out mechanical flexure as a significant source of calibration error.  Exposures were also taken in the same band over several runs to test for changes in the flat-field due to instrument movements.  No significant changes are seen in the flat-field between runs spaced as much as a year apart.  In contrast, the scattered light pattern, and hence the dome flat, changed by as much as 2\%.   
	
	Since no variations were observed in the scattered light corrected flat-field as a function of telescope position, time, or photometric quality, the same flat can be and was used for multiple telescope runs.  These flats are publicly available as part of the COSMOS archive.  An example of the correction for the $r^+$ band is shown in Figure \ref{f:dome-flat-corr}.
			
	We find that the scattered light component is largest in the outer $8\arcmin$ of the field of view which is vignetted by the primary mirror.   In this outer region the typical correction to a dome flattened image is 2-4\%.  In contrast, the central $26\amin$ of the field-of-view is stable and flat to 1\% with no calibration.  Therefore the scattered light can be safely ignored for the inner regions of Suprime-Cam. 
	
\subsubsection{Photometric Calibration and Image Combination}

	After flat fielding object frames taken on photometric nights are corrected for atmospheric extinction measured from standard stars \citep{aussel-photom}.  Data taken on non-photometric nights are scaled to those taken in photometric conditions using the $P_e$ factors calculated during the scattered light correction.  Exposures with extinction greater than 0.5 magnitudes are discarded.  Absolute photometric calibration is done on the AB system using the Subaru filter transmission curves.  As a result, all images are in units of nJy per pixel.  Color conversion and methodology are discussed in \citet{aussel-photom}.

	After calibration the images in any given band are smoothed to the same full-width at half maximum (FWHM) using a Gaussian kernel.  They are then resampled onto the final astrometric grid with a sum-over-triangles interpolation using the IMCAT `warpimage' task.  Inverse variance-maps, derived from the image noise and flat-fields, are also generated and resampled on to the final astrometric grid.  The variance is scaled so that the noise measured in a given sky area on the resampled images is identical to that measured in the same area on the original images.
	
 	Once re-sampled, the images and variance maps are combined with the IMCAT  `combineimages' command using a weighted sum with outlier pixels, more than $5\sigma$ from the median, removed.  A final RMS map is also generated by `combineimages' that reflects the true pixel-pixel RMS.  The PSF homogenized images provide the most consistent photometry but lose some sensitivity due to the smoothing. 
	
	A second combination was done with the original PSF images for all bands to provide a maximum sensitivity image for detection.  Since the PSF varies as a function of position and magnitude in these images the aperture photometry and colors are less reliable.  Finally, for the $B_J$, $r^+$, and $i^+$ data a third combination of the images was done with only the best seeing data.  The resulting images have a FWHM of 0.6\asec, 0.8\asec, and 0.5\asec respectively. 
	
\subsection{CFHT Megaprime }\label{s:CFHT-redux}

	The Megaprime camera has a 1 sq degree field of view on the 3.6m CFHT  telescope.  The focal plane is covered with 36 2k $\times$ 4.5k EEV CCD detectors with excellent response between 3200\AA\ and 9000\AA\ \citep{megaprime,2003SPIE.4841...72B}.  	
		
\begin{deluxetable}{cccc}
\tabletypesize{\scriptsize}
\tablecaption{Summary observing log for CFHT \label{t:megaprime-obs-log}}
\tablehead{
\colhead{Telescope}&\colhead{Filter}&\colhead{Exposure}&\colhead{Observation}\\
\colhead{}			&\colhead{}	  &\colhead{Time (h)}&\colhead{Date} 
}
\startdata
CFHT	&	$u^*$	&	1.8	 &	2003-12-21\\
		&			&	1.8 	&	2003-12-22\\
		&			&	1.9	&	2004-01-19\\
		&			&	3.4 	&	2004-01-20\\			
		&			&	1.8 	&	2004-04-22\\
		&			&	1.8 	&	2004-04-25\\
		&			&	0.6	&	2004-05-22\\
		&			&	0.3 	&	2005-04-04\\
		&			&	0.2	&	2005-04-05\\
		&			&	1.5	&	2005-04-09\\	
		&			&	1.0 	&	2005-04-11\\
		&			&	1.0	&	2005-04-14\\
		&			&	1.0	&	2005-05-04\\
		&			&	0.9	&	2005-05-05\\
		&			&	0.9	&	2005-05-06\\
		&			&	0.9	&	2005-05-08\\
		&			&	0.4	&	2005-05-29\\
		&			&	0.4 	&	2005-06-02\\
		&			&	0.2 	&	2005-06-03\\
		&			&	0.4 	&	2005-06-04\\
		&			&	0.4 	&	2005-06-05\\
		&			&	0.2 	&	2005-06-06\\
CFHT       &	$i^*$		&	0.1	&	2003-12-21\\
		&			&	0.9 	&	2004-01-15\\
		&			&	1.3 	&	2004-01-17\\
		&			&	2.9 	&	2004-01-18\\
\enddata
\end{deluxetable}

	   Megaprime was used to obtain deep u$^*$ band (3798\AA) and shallow i$^*$ band images of the COSMOS field.  Objects as bright as 15$^{th}$ magnitude are unsaturated in long exposures with Megaprime, allowing for an excellent astrometric solution.   Observations were taken in sets of five dithered exposures forming a five-point dice-face pattern.  A total of five overlapping pointing centers were observed to cover the COSMOS field, resulting in data four times deeper in the center of the field than on the edges.  The depth variation is recorded in the noise maps discussed in Section \ref{s:data-products}.  A summary of the observing logs for the Megaprime observations is given in Table \ref{t:megaprime-obs-log}.
	   
	   The CFHT operates in a queue observing mode for Megaprime observations which ensures uniform image quality and photometry.  CFHT and TERAPIX also provide a standard reduction pipeline which meets our calibration requirements. 
	   	  	
	Appropriate calibration frames were taken each night by the queue observer, and calibrated data were provided by the Elixir pipeline \citep{elixir}.  This pipeline corrects for bias, dark current, flat fielding, and scattered light, with the final photometric calibration better than 1\% across the field of view.  
   	
	Further reduction including astrometric and photometric calibration, sky subtraction, and image combination was provided by the TERAPIX data processing center\footnote{\url[http://terapix.iap.fr/soft/]{http://terapix.iap.fr/soft/}}.  At TERAPIX the calibrated images provided by Elixir were visually inspected with the `qualityFITS' data quality assessment tool and any defective images were rejected.  Images with a seeing larger than 1.3\asec in the $i^*$ band and 1.4\asec in the $u^*$ band were also rejected.  

	All of the images were astrometrically registered to the COSMOS catalog \citep{aussel-photom} using the `Astrometrix' package. They were then resampled using a Lancsos-3 interpolation kernel and median-combined using the `SWarp' image combination software.  Although the median combination is sub-optimal in signal-to-noise, it provides the best rejection of cosmic rays.  RMS-maps, derived from the image noise and flat-fields, were combined using the same astrometric solution. The image scale for the final stack was set to 0.15\asec per pixel.  
	
	The $u^*$ band images were processed twelve months after the $i^*$ images.  The reduction was similar, except a new TERAPIX tool `Scamp'  was used to compute the global astrometric and photometric solutions for the $u^*$ images.   Finally, a range of quality assessment tests were conducted on the final images similar to those described in \citet{2003A&A...410...17M}.  
			
\subsection{CTIO ISPI and KPNO FLAMINGOS }\label{s:IR-redux}

	The Flamingos camera on the KPNO 4m \citep{flamingos} and the ISPI camera on the CTIO 4m \citep{ISPI} both provide a field of view slightly larger than $10\amin \times 10\amin$.  These cameras contain a single 2k $\times$ 2k HAWAII-2 infrared array detector which is sensitive between $0.9\mu m$  and $2.4\mu m$.   Data from these instruments were combined to obtain a $K_s$ band image covering the entire COSMOS field.  However, due to weather and the instrument field of view, the depth varies with position.  The average KPNO exposure time is 1596 sec in $K_s$ band over the field.  With CTIO an exposure time of 1436 sec in $K_s$ was obtained over the whole field.  The variation in depth is recorded in the noise map discussed in Section \ref{s:data-products}.  A summary of the observing log for the KPNO/CTIO observations is given in Table \ref{t:ir-obs-log}.
	
	Eighty-one KPNO or CTIO pointings were required to cover the entire COSMOS field.  Every position was covered at least four times with KPNO and three times with CTIO.  A second grid of 64 pointings offset by half a pointing was taken with CTIO to ensure photometric consistency.  The central nine pointings of the COSMOS field were covered with additional passes at the end of the 2004 KPNO run.  At each pointing a rotated five-point dice-face dither pattern with a 1\arcmin\ diameter was used.  The central position was offset randomly by a few arc-seconds between passes to reduce the likelihood of bad pixels falling on the same portion of the sky.

\begin{deluxetable}{cccc}
\tabletypesize{\scriptsize}
\tablecaption{Summary observing log for CTIO and KPNO\label{t:ir-obs-log}}
\tablehead{
\colhead{Telescope}&\colhead{Filter}&\colhead{Exposure}&\colhead{Observation}\\
\colhead{}			&\colhead{}	  &\colhead{Time (h)}&\colhead{Date} }

\startdata
KPNO	&	$K_s$	&		3.6	& 2004-02-05\\
		&	$K_s$	&		4.3	& 2004-02-06\\
		&	$K_s$	&		5.9	& 2004-02-07\\
		&	$K_s$	&		5.0	& 2004-02-08\\
		&	$K_s$	&		5.2	& 2004-02-09\\
			&	$K_s$		   &			   3.8	& 2005-03-31\\
			&	$K_s$		   &			   3.5	& 2005-04-01\\
			&	$K_s$		   &			   5.6	& 2005-04-02\\
CTIO	&	$K_s$	&		0.3	& 2004-04-05\\
		&	$K_s$	&		3.2	& 2004-04-06\\
		&	$K_s$	&		3.8	& 2004-04-07\\
		&	$K_s$	&		3.8 	& 2004-04-08\\
		&	$K_s$	&		3.7	& 2004-04-09\\
		&	$K_s$	&		3.2	& 2004-04-10\\
		&	$K_s$	&		2.4	& 2005-03-29\\
		&	$K_s$	&		3.4	& 2005-03-30\\
		&	$K_s$	&		2.8	& 2005-03-31\\
		&	$K_s$	&		2.8	& 2005-04-01\\
		&	$K_s$	&		3.0	& 2005-04-02\\						
		
\enddata
\end{deluxetable}

	The data were reduced using IRAF\footnote{\url[http://iraf.noao.edu/]{http://iraf.noao.edu/}} with a full double-pass reduction algorithm.  In the first pass sky flats are produced from averaged, sigma clipped, subsets of 20-30 dark-subtracted images with similar sky levels.  A cross-check of different sky flats throughout the night shows only low level, large scale, slowly changing gradients.  A global bad pixel mask is generated from a flat used to identify the dead pixels and a pair of dark exposures used to identify hot pixels.  The science data were then dark subtracted and flat-fielded.  Accurate positional offsets were determined using IRAF's `Imalign' task using multiple well-detected sources identified by SExtractor \citep{1996A&AS..117..393B} that were common to each dithered data set.  The images were then stacked using integer pixel offsets with IRAF's imcombine task.  These initial stacks of the science data were used to generate relatively deep object masks through SExtractor's `CHECKIMAGE\_TYPE = OBJECTS' output file.  These object masks were used to explicitly mask objects when re-generating the sky flats and in the second pass reduction.  Supplementary masks for individual images were made to mask out satellites or other bad regions not included in the global bad pixel mask on a frame-by-frame basis. 

In the full second pass reduction algorithm we individually subtracted the sky from each science frame with 8-10 temporally adjacent images.  Each sky-subtracted image was then flat-fielded with the corresponding object-masked sky flat and any residual variations in the sky removed by subtracting a constant to yield a zero mean sky level.  These individual sky-subtracted images were then masked using a combination of the object mask and any supplementary mask to remove the real sources and bad regions in the sky frames and averaged with sigma-clipping to remove cosmic rays.   These images were further cleaned of any non-constant residual gradients as needed by fitting to the fully masked (object $+$ supplementary $+$ global bad pixel masks) background on a line-by-line basis.  The dithered data sets were re-stacked with the same offsets determined in the first pass using the global and supplementary masks.  Finally, an initial astrometric solution was determined using the 15-50 stars on each frame from the USNO-A2 catalog \citep{1998yCat.1252....0M}.

\begin{figure*}
\begin{center}
\begin{tabular}{ccc}
\includegraphics[scale=0.3]{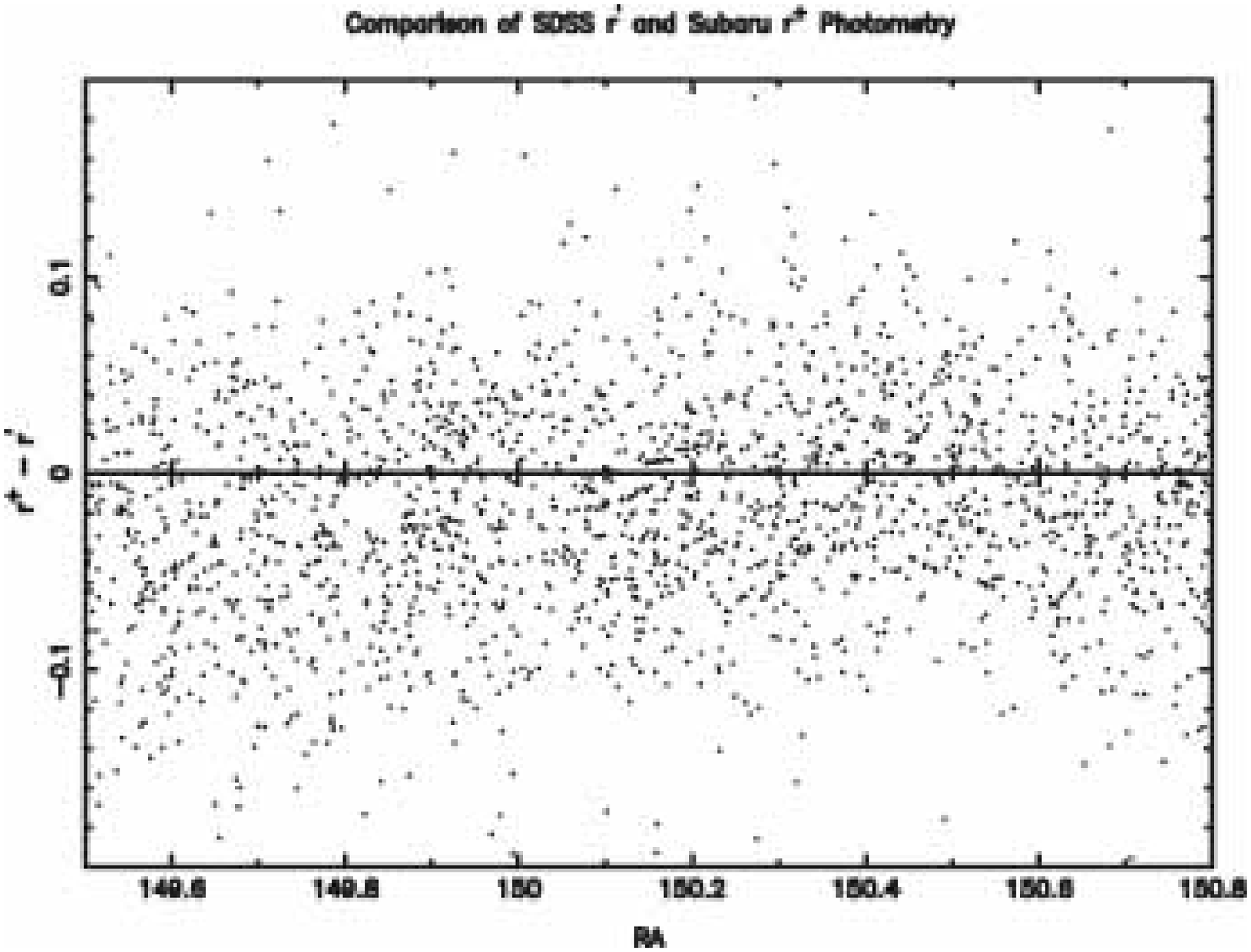}& &\includegraphics[scale=0.3]{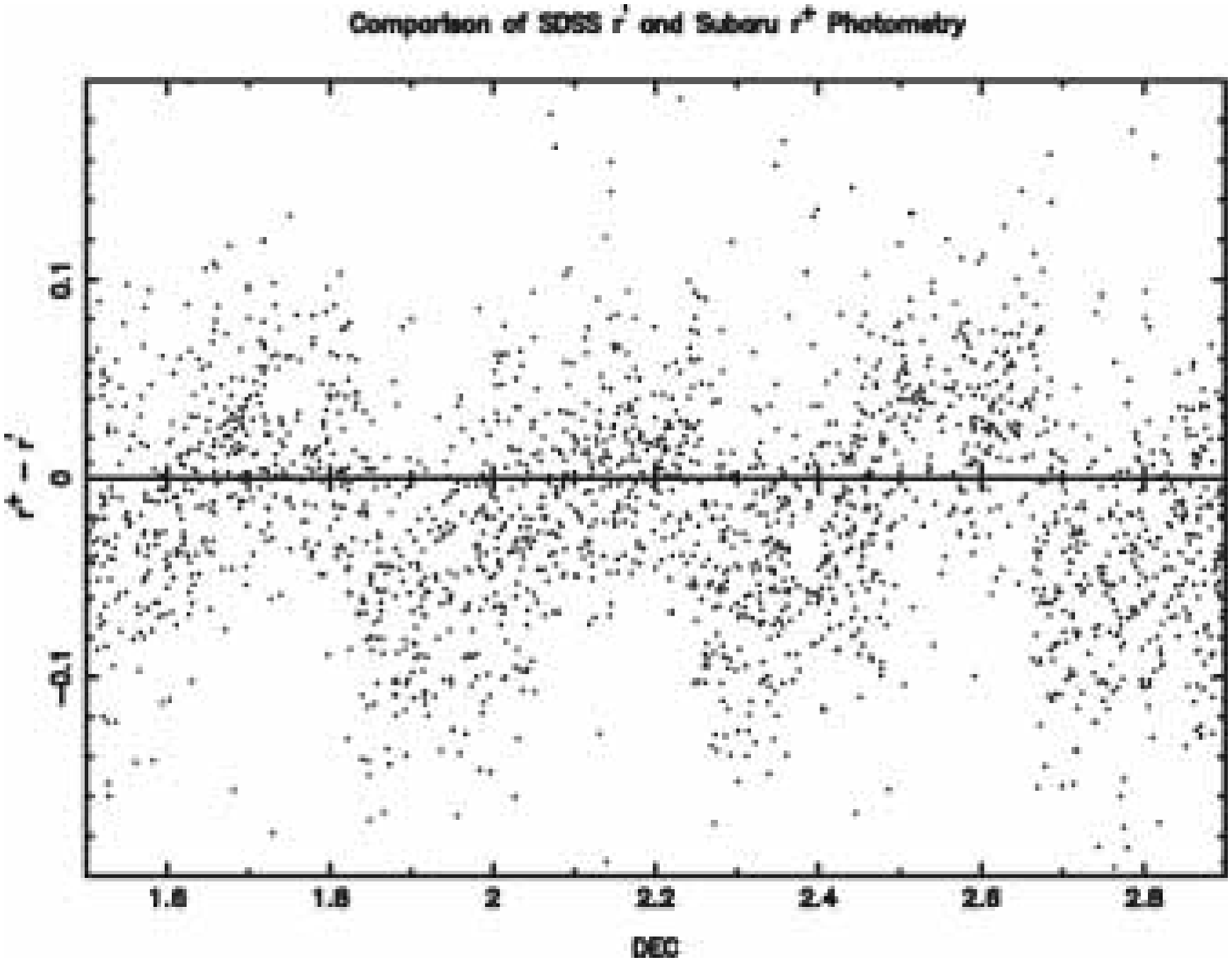}\\
\end{tabular}
\caption{Photometric offsets for stars are shown between the SDSS $r$ and Subaru $r^+$ data as a function of RA (left) and DEC (right).  Notice there are no systematics in RA, but there is a step like pattern in DEC which corresponds to two different SDSS runs.  The step pattern is due to imperfect PSF matching. \label{f:sdss-offset}}
\end{center}
\end{figure*}

	After flat-fielding and sky subtraction, frames with seeing worse than 1.5\asec were removed and all remaining frames were visually inspected.  Any frame with especially high noise, poor tracking, or other defects was removed.  Next, every frame was registered to the COSMOS astrometric catalog with a 4th order, 2-dimensional polynomial.  The polynomial fits were improved by removing mismatched objects in an iterative fashion until the solution converged (typically in 4 iterations).  The resulting scatter between the fit positions and the final astrometry was always less than 0.3\asec\ independent of position.  The large astrometric scatter was due to the 0.3\asec  pixel size of the detector and poor seeing.
		
	Each image was then scaled so that its photometry agrees with the 2MASS point source catalog \citep{2006AJ....131.1163S}.  A shift of 1.852 magnitudes was applied to the 2MASS magnitudes to convert them to the AB system.  The 2MASS, CTIO, and KPNO $K_s$ filters are sufficiently similar so that no color terms are needed to convert between these filters.  After photometric calibration, the scatter between overlapping exposures is measured to remove any position dependent photometric shifts.  
	
	Finally, all data were smoothed with a Gaussian kernel to a 1.5\asec\ FWHM and combined with `SWarp', to produce images on the COSMOS grid with 0.15\asec\ per pixel.  RMS noise maps, derived from the image noise and flat fields, were also combined using the same astrometric solution.  An original PSF image was not produced for the $K_s$ band data because the PSF variations were too large.

\subsection{Sloan Digital Sky Survey }\label{s:SDSS-redux}

	SDSS second data release (DR2) images \citep{SDSS-DR2}  were used to obtain photometry for objects saturated in the Subaru and CFHT data.   Objects as bright as 10$^{th}$ magnitude have good photometry in the SDSS images.  DR2 was used because later SDSS data releases contained no new data or calibration for the COSMOS field.
	
	To facilitate photometric measurements, a mosaic of the SDSS data was created on the same grid as the other COSMOS data.  The data were downloaded and calibrated with the "Best" photometric calibration as outlined on the SDSS-DR2 web site \footnote{\url[http://www.sdss.org/dr2/]{http://www.sdss.org/dr2/}}.    A median sky level was then subtracted and a catalog generated for each image.  The objects in each image were then matched to the COSMOS astrometric catalog using a 3$^{rd}$ order 2-dimensional polynomial with the IMCAT `fitgeometry' routine. 
	
		The PSF of the SDSS data was homogenized by smoothing all images to the same FWHM with a Gaussian kernel.  However, systematic effects of the order of 5\% remain between SDSS `stripes' due to non-Gaussian wings of the SDSS PSF.  This is discussed further in Section \ref{s:psf-match}.  The COSMOS archive also contains a second combination with the un-PSF homogenized images. 
		
		After PSF homogenization, images were re-sampled with a linear interpolation using the IMCAT `warpimage' routine.  All images in each band were then combined with the IMCAT `combineimages' routine using a weighted average.  A RMS noise map was also generated during the image combination process.  The image scale for the final stack was set to 0.15\asec per pixel.

\section{Point Spread Function (PSF) Matching }\label{s:psf-match}

\begin{deluxetable}{lccc}
\tabletypesize{\scriptsize}
\tablecaption{PSF properties and smoothing kernels used for photometry\label{t:smkern}}
\tablehead{
\colhead{Filter} & \colhead{PSF FWHM} & \colhead{Fraction of Flux}      & \colhead{Smoothing Kernel}\\
\colhead{}          & \colhead{(\asec)}          & \colhead{in 3\asec Aperture} & \colhead{$\sigma$ (\asec)\tablenotemark{*}}\\
}
\startdata
$u$&	1.97	&	0.716	&	0.000\\	
$u^*$	&	0.90	&	0.919	&	0.662\\
$B_J$	&	0.95	&	0.942	&	0.699\\
$g$&	1.97	&	0.725	&	0.000\\
$g^+$	&	1.58	&	0.795	&	0.240\\
$V_J$	&	1.33	&	0.874	&	0.521\\
$r^+$	&	1.05	&	0.914	&	0.639\\
$r$&	1.97	&	0.708	&	0.000\\
$i$&	1.97	&	0.709	&	0.000\\
$i^+$	&	0.95	&	0.914	&	0.611\\
$i^*$		&	0.95	&	0.891	&	0.620\\
$NB816$	&	1.51	&	0.851	&	0.463\\
$F814W$	&	0.07	&	0.979	&	0.785\\
$z$&	1.97	&	0.701	&	0.000\\
$z^+$	&	1.15	&	0.866	&	0.585\\
$K_s$	&	1.50	&	0.759	&	0.000\\
\enddata
\tablenotetext{*}{FWHM $=2\sqrt{2 ln(2)}\sigma$}
\end{deluxetable}

\begin{deluxetable*}{lll}
\tabletypesize{\scriptsize}
\tablecaption{SExtractor Parameters \label{t:sex-param}}
\tablehead{
\colhead{Parameter} & \colhead{Setting} & \colhead{Comment}}
\startdata
PARAMETERS\_NAME 	& cosmos.param		& Fields to be included in output catalog\\
FILTER\_NAME		& gauss\_2.5\_5x5.conv	& Filter for detection image\\
STARNNW\_NAME    	& default.nnw                	& Neural-Network\_Weight table filename \\
CATALOG\_NAME    	& STDOUT                       	& Output to pipe instead of file \\
CATALOG\_TYPE    		& ASCII                          	& Output type\\
DETECT\_TYPE     		& CCD                            	& Detector type\\
DETECT\_MINAREA  	& 3                              	& Minimum number of pixels above threshold\\
DETECT\_THRESH   	& 0.6                            	& Detection Threshold in $\sigma$\\
ANALYSIS\_THRESH 	& 0.6                            	& Limit for isophotal analysis $\sigma$\\
FILTER          			& Y                              	& Use filtering \\
DEBLEND\_NTHRESH 	& 64                             	& Number of deblending sub-thresholds\\
DEBLEND\_MINCONT 	& 0.0                            	& Minimum contrast parameter for deblending\\
CLEAN           			& Y                              	& Clean spurious detections\\
CLEAN\_PARAM     		& 1                              	& Cleaning efficiency\\
MASK\_TYPE       		& CORRECT                     & Correct flux for blended objects\\
PHOT\_APERTURES  	& 6.7,13.3,20,26.7,33.3,66.7 & MAG\_APER aperture diameter(s) in pixels\\
PHOT\_AUTOPARAMS 	& 2.5, 3.5                       	& MAG\_AUTO parameters: $<$Kron\_fact$>$,$<$min\_radius$>$\\
PHOT\_FLUXFRAC   	& 0.2,0.5,0.8,0.9                & Define n-light radii\\
PHOT\_AUTOAPERS  	& 20.0, 20.0                     	& MAG\_AUTO minimum apertures: estimation, photometry\\
SATUR\_LEVEL     		& 200000                         	& Level of saturation\\
MAG\_ZEROPOINT   	& 31.4                           	& Magnitude zero-point\\
GAIN            			& 1					& Gain is 1 for absolute RMS map\\
PIXEL\_SCALE     		& 0.1500                         	& Size of pixel in \asec\\
SEEING\_FWHM     		& 0.95                           	& Stellar FWHM in \asec\\
BACK\_SIZE       		& 256                            	& Background mesh in pixels \\
BACK\_FILTERSIZE 		& 5                              	& Background filter\\
BACKPHOTO\_TYPE  	& LOCAL                          	& Photometry background subtraction type\\
BACKPHOTO\_THICK 	& 120                            	& Thickness of the background LOCAL annulus\\
WEIGHT\_GAIN 		& N                                  	& Gain does not vary with changes in RMS noise\\
WEIGHT\_TYPE 		& MAP\_RMS                     & Set Weight image type\\
MEMORY\_PIXSTACK 	& 5000000                        	& Number of pixels in stack\\
MEMORY\_BUFSIZE  	& 4096                           	& Number of lines in buffer\\
MEMORY\_OBJSTACK 	& 60000                          	& Size of the buffer containing objects\\
VERBOSE\_TYPE    		& QUIET                        	& \\
INTERP\_MAXXLAG   	& 3					& Number of bad pixels to interpolate over in X\\
INTERP\_MAXYLAG   	& 3					& Number of bad pixels to interpolate over in Y\\
INTERP\_TYPE      		& ALL				& Type of Interpolation\\
\enddata
\end{deluxetable*}
	
	A consistent PSF within each band and between bands is essential for high-quality photometry. Ideally, all bands would have an identical point spread function (PSF), but achieving a homogeneous PSF for a data set as diverse as COSMOS is extremely difficult due to the non-gaussian portion of most PSFs.  To ensure a consistent PSF across the field of view and between bands we adopted a two step process.  First we homogenize the PSF within each band during the data  reduction, then we match the homogenized PSFs to the band with the worst image quality.

	Within each band, we adopt a Gaussian kernel to homogenize the PSF between exposures.  This works well if the seeing variations are small, there are many images at the same position to average out the PSF, and the photometric apertures are much larger than the seeing size.  These assumptions, however, break down for large seeing variations and small numbers of images.  In the COSMOS data the effect of non-Gaussian PSF components is negligible ($\le0.01$ magnitudes) for all but the SDSS data when using the PSF matched 3\asec aperture photometry. 

	The SDSS data consist of a single exposure at each position with seeing variations as large as 1\asec between the two nights which cover the COSMOS field (see Section \ref{s:data-redux} and \ref{s:SDSS-redux}).  These data are collected in five parallel strips in RA with a detector wide gap in declination between strips.  Two passes are required to completely cover an area of the sky.  Figure \ref{f:sdss-offset} shows the offset between the SDSS $r$ and the Subaru $r^+$ band photometry measured in a 3\asec aperture as a function of Right Ascension and Declination.  An offset of 0.06 magnitudes is clearly seen in declination between the two SDSS passes due to the imperfect PSF matching.  This offset is not visible if total magnitudes, which correct for seeing variation, are compared (see Section \ref{s:photom-consistency}).  The offsets are similar in all SDSS bands, so colors between SDSS bands are not significantly impacted.

	For stellar photometry measured with the Subaru data in small ($\simeq1$\asec) apertures a magnitude-dependent aperture correction is required between 18th and 21st magnitude.   The correction is due to differences in seeing between the short and long exposure data taken with Suprime-Cam.  The short exposure data were typically taken at higher air mass, and hence worse seeing, than the long exposure data.  In addition, fewer exposures were taken with shorter than longer exposure times.  As a result, the long exposure data are smoothed with a much larger kernel than the short exposure data resulting in better PSF matching at fainter magnitudes.  Corrections for this effect are given for several bands in \citet{rich-stars}, which uses a different photometric catalog from the one presented here. 

 	To avoid these PSF matching problems in the multi-band catalog, the PSF matching was optimized for a 3\asec aperture.  This was achieved by convolving each PSF homogenized image with a Gaussian kernel that produced the same flux ratio between a 3\asec\ and 10\asec\ aperture for a point source.  This method is superior to simply matching the FWHM since it accounts for the non-Gaussian parts of the PSF.  However, with this method, only 3\asec apertures are free from systematic effects. 
	
	We selected the kernel for each image by identifying point sources in the ACS images, then using these to construct a median PSF for each of the PSF homogenized images.  A Gaussian kernel was then selected which yielded the same flux ratio between a 3\asec and 10\asec aperture as the band with the smallest ratio (the $K_s$ image).  These smoothing kernels are listed in Table \ref{t:smkern}.
	
	 With the exception of the SDSS images, the worst seeing image was the $K_s$ band which contained only 76\% of the flux in a 3\asec aperture.    To account for the fact that the SDSS images have seeing worse than the $K_s$ image, an aperture correction of -0.06 magnitudes was applied to the photometry from these bands.  

\section{The Multi-Band Catalog}\label{s:catalog}

	The COSMOS multi-band catalog is derived from a combination of the the CFHT $i^*$ and Subaru $i^+$ original-PSF images.  The CFHT $i^*$ band image alone is too shallow, while compact objects in the Subaru $i^+$ image saturate at $21^{st}$ magnitude.  Therefore, a combination of the two gives the largest possible dynamic range for a detection image.   The resulting catalog is well matched in wavelength and depth to the HST-ACS catalog \citep{acs-weak-lensing} and optimally de-blends the ground based photometry.  The catalog is also optimal for many science goals which require photometric redshifts.
		
	   A $\chi^2$ image constructed from multiple bands was also tried as a detection image but then rejected.  The main advantage of a $\chi^2$ image is pan-chromatic completeness.  However, the depth and quality of the $i^+$ band image mean only faint objects with very extreme colors would be detected in a $\chi^2$ image but not the $i^+$ band image.  As a result of their faint magnitudes and extreme colors these objects will have poor photometric redshfits, so detecting them is of limited use.  Furthermore, the scientific drivers for a $\chi^2$ image strongly favor including the $K_s$ band which has 1.5\asec seeing.  This poor seeing significantly reduces the ability to split close pairs of objects, so the benefits of a $\chi^2$ image are outweighed by the loss in resolution.
	  
	The catalog was generated with SExtractor \citep{1996A&AS..117..393B} run in dual image mode on each of the 144 image tiles (see Section \ref{s:data-products}).  Only objects in the central $10 \amin \times 10\amin$ of each tile were recorded to avoid duplicate detections in the overlapping regions.  The combined $i^+$ and $i^*$ band image was used as the detection image for all bands.  Absolute RMS maps were used as weight maps for both the detection and measurement images\footnote{The default SExtractor settings assume variations in noise reflect variations in gain across the image.  However, in this case both the data and RMS images are in absolute flux units, so the gain is 1 everywhere by definition.  The SExtractor parameter `WEIGHT\_GAIN' must be set to N to avoid the default behavior.  Failure to set `WEIGHT\_GAIN=N' will result in incorrect error estimates.}.  Measurements were done on the PSF homogenized images, further smoothed to PSF match all bands (see Section \ref{s:psf-match}).
	
	The number of contiguous pixels, detection thresholds, de-blending parameters, and smoothing kernels were adjusted to maximize completeness when visually compared to the ACS and $\chi^2$ image.  Setting these parameters aggressively results in false detections around bright stars and residual image defects.  However, these false detections can be removed by requiring $5\sigma$ measurements in the detection band,  a reasonable FWHM, and a defect mask for the detection image.   The SExtractor parameters are given in Table \ref{t:sex-param}, while the software used to generate these catalogs is available at the COSMOS website \footnote{\url[http://cosmos.astro.caltech.edu]{http://cosmos.astro.caltech.edu}}.

	Aperture photometry with a 3\asec diameter is measured for each band.  This provides the best possible color measurements by minimizing the effects of PSF variation from band to band (see Section \ref{s:psf-match}).  Total magnitudes are sub-optimal for color measurements because a correction factor must be estimated separately for each band, increasing the error.  Furthermore, SExtractor estimates this correction only on the detection image, so estimates of total magnitude are only accurate if the image quality of the detection and measurement image are identical.
		
	In contrast, properly PSF matched images have identical aperture corrections in all bands, so a single estimate of the total magnitude can be used for all bands.  We estimate the correction to total magnitude using the offset between the total (MAG\_AUTO) and the 3\asec\ aperture magnitude in the detection image.  This difference can be added to any aperture magnitude to yield a total magnitude.

\begin{deluxetable}{lcccc}
\tabletypesize{\scriptsize}
\tablecaption{Depth of $i$ Band Catalog Photometry\label{t:cat-depth}}
\tablehead{
\colhead{Filter} & \colhead{$5\sigma$ Depth} 		& \colhead{RMS Range}  & \colhead{Upper\tablenotemark{1}} & \colhead {Lower\tablenotemark{2}} \\
\colhead{}          & \colhead{in 3\asec Aperture}          & \colhead{of Depth}        &  \colhead{Quartile} & \colhead{Quartile}\\
}
\startdata
$u$&	23.2 & 0.3  & 22.9 & 23.4\\	
$u^*$	&	26.5 & 0.2 & 26.3 & 26.6\\
$B_J$	&	26.6 & 0.1 & 26.6 & 26.7\\
$g$&	23.9 & 0.2 & 23.9 & 23.9\\
$g^+$	&	26.5 & 0.2 & 26.5 & 26.6\\
$V_J$	&	26.5 & 0.2 & 26.4 & 26.5\\
$r^+$	&	26.6 & 0.2 & 26.5 & 26.6\\
$r$&	23.6 & 0.3 & 23.4 & 23.8\\
$i$&	22.9 & 0.3 & 22.8 & 23.2\\
$i^+$	&	26.1 & 0.2 & 26.0 & 26.2\\
$i^*$		&	23.5 & 0.3 & 23.3 & 23.7\\
$NB816$	&	25.5 & 0.2 & 25.4 & 25.6\\
$F814W$	&	25.3 & 0.1 & 25.3 & 25.4\\
$z$&	21.5 & 0.3 & 21.3 & 21.7\\
$z^+$	&	25.1 & 0.2 & 25.1 & 25.2\\
$K_s$	&	21.2 & 0.3 & 21.1 & 21.3\\
\enddata
\tablenotetext{1}{25\% of objects with $5\sigma$ measurements are brighter than this magnitude}
\tablenotetext{2}{25\% of objects with $5\sigma$ measurements are fainter than this magnitude}
\end{deluxetable}

	 The median $5\sigma$ depths for the catalog including PSF matching, de-blending, background subtraction, and photon noise are given in Table \ref{t:cat-depth}.  These numbers are for total magnitudes and should be used when choosing signal to noise cuts and magnitude limits for the COSMOS catalog.  An estimate of the RMS variation in the $5\sigma$ limiting magnitude, along with the upper and lower quartiles of the $5\sigma$ limiting magnitude is also given.
	 	 
	The general release catalog is cut at a total $i^+$ magnitude of 25 and only includes the 2 square degrees with uniform multi-band coverage.  At this magnitude limit and in this area photometric redshifts are reliable, the catalog is complete, and spurious sources are minimal.  At fainter magnitudes the catalog begins to be incomplete, have more spurious detections, and photometric redshifts begin to behave poorly.  A full catalog of all detections is available on request. However, the full catalog should be used with caution.  In particular the completeness and the number of spurious sources will vary as a function of position due to differences in the RMS background noise.

\subsection{Catalog Contents}\label{s:cat-contents}

	The multi-band catalog contains PSF matched 3\asec\ aperture photometry and errors for all Subaru and CFHT bands along with the KPNO+CTIO $K_s$ band and HST F814W band data.  Non-PSF matched photometry is also included for the SDSS bands.  Objects with no detection are assigned a magnitude of 99 and an error indicating the $1\sigma$ limiting magnitude expected for that band.  Objects with no measurement due to lack of coverage, saturation, or other defects are assigned a magnitude and error of -99.
	
		The catalog photometry uses the photometric zero points determined by the standard stars, which are known to have systematic offsets on the order of 0.05 magnitudes and up to 0.2 magnitudes in $B_J$ band.  We strongly suggest applying the zero point corrections given in Table \ref{t:zp-corr} and discussed in Section \ref{s:zp-corr} to get the best possible photometry. 
	
	The total (MAG\_AUTO) magnitude and the FWHM is measured from either the PSF matched Subaru $i^+$ or the PSF matched CFHT $i^*$ band image (if the Subaru data is missing or saturated) is included in the catalog along with a flag indicating which image was used.  As discussed in the previous section, since the photometry is PSF matched the difference between the total and aperture magnitudes in the combined $i$ band data provides an aperture correction for each object.  This aperture correction is also included in the catalog.  Applying this aperture correction to any band will provide a total magnitude in that band.
		
	A flag indicating that the object may be a star instead of a galaxy is also included, however this flag is a qualitative assessment and science relying on accurate star/galaxy separation should perform a more detailed analysis.  Quantitative indicators such as the SExtractor ``CLASS\_STAR" parameter are ineffective at separating stars and galaxies because the FWHM varies as a function of magnitude on the detection image.  Furthermore, we could not use the HST-ACS data to separate point sources because it only covers a fraction of the deep ground based data.  To overcome this limitation we manually defined regions on a plot of detection image FWHM vs. magnitude which contained the majority of ACS point sources (objects with CLASS\_STAR $> 0.9$ in the ACS catalog \citep{acs-weak-lensing}).  Objects falling in this region were then flagged as stars in the ground based catalog.

	Flags marking masked regions in the Subaru $B_J$, $V_J$, $i^+$, and $z^+$ images are included in the catalog.  These masks were generated by visually inspecting images, using the SDSS magnitudes as guide to flag bright stars and estimate the size of masks.  The flag value is the area in square arc-seconds of the photometry aperture which falls inside a masked region.  
	
	Finally a flag indicating heavily de-blended objects is included in the catalog.  The default SExtractor de-blending settings fail to find objects in areas around bright objects due to the high dynamic range of the $i^+$ band detection image.  Unfortunately, de-blending aggressively enough to find faint objects around bright ones results in numerous false detections near the bright objects.  To mitigate this problem objects not detected by the default SExtractor de-blending parameters are flagged.  

\subsection{Catalog Usage Guide}\label{s:cat-guide}

	Objects with all mask and de-blending flags set to 0 have the most reliable photometry.  However, this removes a significant fraction of the survey area and some flags can be safely ignored under certain circumstances.
	
	 When cross correlating with multi-wavelength detections the de-blending and star flags can often be safely ignored.  However, for clustering analysis based on the optical catalog the de-blending flag is very important, especially at faint magnitudes (see \citet{mccracken_clustering}).  Furthermore, the photometry of  faint objects de-blended from extended nearby galaxies is suspect due to color gradients in the nearby object.  
	 
	The photometry masks are specific to the band in which they were measured. But since the same instrument (Suprime-Cam) was used for most of the photometry they can be safely extended to adjacent bands (for instance a combination of the $B_J$ and $V_J$ mask is appropriate for the $g^+$ photometry).  Photometric redshifts are affected by masked photometry in a non-linear fashion, so all photometry masks must be applied to obtain a clean photometric redshift sample.  However, in specific redshift ranges some bands are not as important, so the masking could be relaxed if a spectroscopic control sample is available.
		
\subsection{Completeness and Confusion}

	For deep surveys the ability to detect objects (completeness) and separate superimposed objects (confusion)  are often more important sources of measurement error than the photon noise.  These quantities are more difficult to quantify than the formal noise given in Table \ref{t:data-depth} because they are sensitive to both the data quality and the software used.   
	
	The image quality, or seeing has a much larger impact on completeness and confusion than the measurement error.  For a fixed aperture the measurement error increases linearly with FWHM, while the peak flux decreases as FWHM$^{-1}$, and the confusion increases as FWHM$^2$.  So a factor of two difference in seeing corresponds to a 0.3 magnitude reduction in measurements sensitivity, but a 0.75 magnitude reduction in peak surface brightness, and a factor of four increase in blending.  As a result it is often possible to achieve much greater flux measurement depth than detection depth.  SExtractor uses a peak finding algorithm which is especially sensitive to image quality and software settings.  The threshold, pixel area, and smoothing kernel settings have a large impact on completeness while the deblending parameters impact confusion.  These settings are often a compromise between detection depth and the number of spurious detections.  

\begin{figure}
\begin{center}
\includegraphics[scale=0.3]{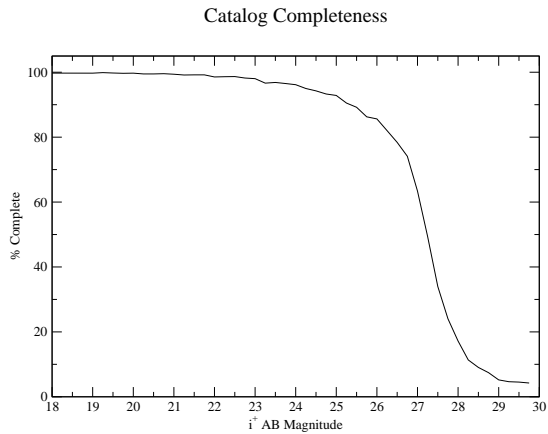}
\caption{The estimated completeness of the $i^+$ band detected catalog is shown as a function of magnitude using the method described in \citet{2003A&A...410...17M}.  This method randomly places objects with a representative range of morphologies in the field, and attempts to recover them.  The estimate includes the effects of blending and detection completeness.\label{f:cat-completeness}}
\end{center}
\end{figure}
	
	For the COSMOS catalog we chose parameters which maximized detection depth but also produced spurious detections near the detection limit.  These spurious objects can be removed with object masks and cuts in magnitude and FWHM.  A similar method to ours is employed by the CFHT-LS survey.  However, other groups such as the Subaru XMM Deep Survey (SXDF), have preferred more conservative settings which minimize spurious sources at the expense of detection sensitivity.  These more conservative numbers are given for all Subaru bands in \citet{tanaguchi-cosmos}.
	
	To quantify completeness, simulated objects with a representative range of morphologies and magnitudes are inserted into the image.  SExtractor is then run on the image, and the fraction of recovered objects at each magnitude is measured.  No attempt is made to avoid existing objects, so the effects of confusion are included in the completeness calculation. This is identical to the method described in \citet{2003A&A...410...17M}.  Figure \ref{f:cat-completeness} shows the results of this simulation for the combined $i^+$ and $i^*$ detection image.  The catalog is 91\% complete at $i^+=25.0$, 87\% complete at $i^+=26.0$ and 50\% complete at $i^+=27.4$. 	
	
\section{Discussion}\label{s:discussion}

\subsection{Consistency of Photometry}\label{s:photom-consistency}

	The COSMOS survey was designed to probe the interplay of large scale structure, and galaxy evolution.  High quality photometric redshifts are essential for these studies.   Uncertainties greater than 2\% in redshift will begin to wash out large scale structure \citep{scoville-hst}, so systematic variations in photo-z's need to be less than 1\%.  Since a 1\% error in photometry typically leads to a 1\% error in photo-z, this places extremely high requirements on the input photometry quality.
	
	The Canada-France-Hawaii Telescope Legacy Survey (CFHT-LS) and the SDSS have independently covered the COSMOS field, allowing for a quantitative estimate of our photometric consistency.  The Subaru $i^+$, CFHT $i^*$, and HST-ACS F814W data have similar band-passes, but were collected on different instruments and reduced by different teams using different software, allowing for an internal check on photometric consistency.
	
	The best effort at estimating total magnitude from each survey is used when comparing data to minimize effects of aperture size.  These specific magnitudes used were SExtractor MAG\_AUTO values for CFHT-LS, Petrosian magnitudes for SDSS, and aperture magnitudes corrected to total (as described in Section \ref{s:catalog}) for COSMOS.  The CFHT-LS and SDSS photometry were then converted to the COSMOS filter system using the conversions given in Tables \ref{t:color-conv-SDSS} \& \ref{t:color-conv-CFHT} and discussed further in \citet{aussel-photom}.  Finally the catalogs were merged by position with COSMOS allowing for a 1\arcsec\ offset in position.  
	
\begin{figure}
\begin{center}
\includegraphics[scale=0.6]{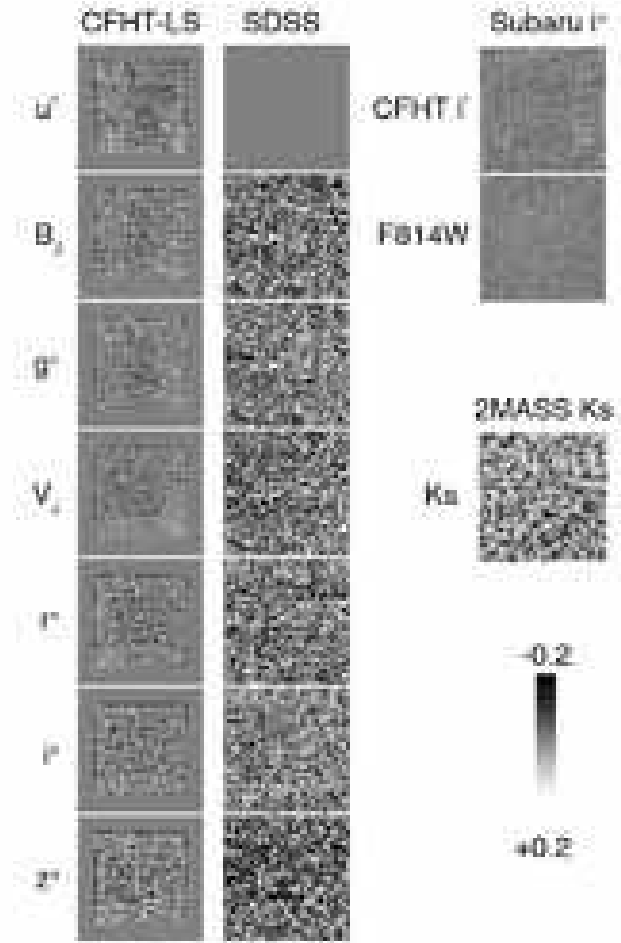}
\caption{The grey scale images show offsets between photometry measured by the CFHT-LS and SDSS surveys and COSMOS (left), between Subaru $i^+$, CFHT $i^*$, and ACS F814W bands within the COSMOS survey (top right), and between 2MASS and COSMOS $K_s$ band.  The images have 1\arcmin\ pixels for all surveys but SDSS and 2MASS, which have 2\arcmin\ pixels.  The CFHT-LS D2 field only covers the central 1 square degree of COSMOS, resulting in a blank area.  A comparison between SDSS $u$ and CFHT $u^*$ is not possible because there is no simple linear relationship between these filters.  The scale is linear in magnitude from black (-0.2 mag) to white (+0.2 mag).  No variation with position is measurable within the measurement errors.  \label{f:phot-consistency} }
\end{center}
\end{figure}

\begin{deluxetable*}{ccl}
\tabletypesize{\scriptsize}
\tablecaption{Color Conversion Between SDSS and COSMOS Photometric Systems\label{t:color-conv-SDSS}}
\tablehead{
\colhead{COSMOS Filter} & \colhead{Valid for } & \colhead{Conversion\tablenotemark{1}}\\
\colhead{}                          & \colhead{}        & \colhead{Equation}
}
\startdata
$B_J$	&	$-1 < (g - r) <  1$ & $B_J = g + 0.240 (g - r) + 0.029$\\ 
$g^+$	&	$-1 < (g - r) <  1$ & $g^+ = g - 0.056 (g - r) - 0.006$\\ 
$V_J$	&	$-1 < (g - r) <  1$ & $V_J = g - 0.617 (g - r) - 0.021$\\ 
$r^+$	&	$-1 < (g - r) <  1$ & $r^+ = r - 0.037 (g - r) + 0.003$\\ 
$i^+$	&	$-1 < (r - i) <  1$ & $i^+ = i - 0.106 (r - i) + 0.007$\\ 
$z^+$	&	$-1 < (i - z) <  0.8$ & $z^+ = z - 0.110 (i - z) + 0.008$\\ 
\enddata
\tablenotetext{1}{A comparison between SDSS $u$ and CFHT $u^*$ is not possible because there is no simple linear relationship between these filters.}
\end{deluxetable*}

\begin{deluxetable*}{ccl}
\tabletypesize{\scriptsize}
\tablecaption{Color Conversion Between CFHT-LS and COSMOS Photometric Systems\label{t:color-conv-CFHT}}
\tablehead{
\colhead{COSMOS Filter} & \colhead{Valid for } & \colhead{Conversion}\\
\colhead{}                          & \colhead{}        & \colhead{Equation}
}
\startdata
$u^*$	& all	&	Same Filter\\
$B_J$	& $-0.2 < (g^* - r^*) < 0.6$ & $B_J = g^* + 0.432 (g^* - r^*) + 0.047$\\
$g^+$	& $-1.0 < (g^* - r^*) < 0.6$ & $g^+ = g^* + 0.094 (g^* - r^*) + 0.008$\\
$V_J$	& $-0.2 < (g^* - r^*) < 0.6$ & $V_J = g^* - 0.545 (g^* - r^*) - 0.016$\\
$r^+$	& $-1.0 < (g^* - r^*) < 0.9$ & $r^+ = r^* - 0.021 (g^* - r^*) + 0.001$\\
$i^+$	& $-1.0 < (g^* - r^*) < 0.8$ & $i^+ = i^* - 0.020 (g^* - r^*) + 0.005$\\
$F814W$	& $0.5 < (i^* - z^*) < 0.0$ & $F814W = z^* + 0.632 (i^* - z^*) - 0.116 (i^* - z^*)^2 - 0.001$\\
$z^+$	& $-1.0 < (i^* - z^*) < 0.8$ & $z^+ = z^* - 0.128 (i^* - z^*) - 0.004$\\
\enddata
\end{deluxetable*}

	Figure \ref{f:phot-consistency} shows a comparison between the COSMOS data and those from the CFHT-LS and SDSS as well as between the Subaru $i^+$, CFHT $i^*$, and HST-ACS F814W bands for objects brighter than 25th magnitude in the COSMOS ground based data and magnitude errors smaller than 0.21 ($5\sigma$) in the comparison data. A comparison between SDSS $u$ and CFHT $u^*$ is not possible because there is no simple linear relationship between these filters. No systematic effects as a function of position are measurable within the photometric uncertainty for the 1\arcmin\ bins used for CFHT-LS and 2\arcmin\ bins used for SDSS.  Smoothing on the size scale of a Suprime-Cam pointing yields a typical RMS variation of 0.01 magnitudes across the field between COSMOS, the CFHT-LS, and SDSS (see Table \ref{t:zpvar}).  While a comparison between the COSMOS CFHT $i^*$ and Subaru $i^+$ gives 0.007 magnitudes of scatter ($1\sigma$), and the ACS-F814W data and Subaru $i^+$ give a dispersion of 0.003 magnitudes.

	Figure \ref{f:phot-consistency} also compares the CTIO and KPNO $K_s$ band photometry to those from 2MASS.  Objects with total magnitudes of $i^+>16$ and at least a $5\sigma$ detection in $K_s$ were used in the comparison.  No systematic trend is measurable with position and the rms variation between COSMOS and 2MASS is 0.02 magnitudes on the scale of a CTIO/KPNO pointing.  At magnitudes brighter than $i^+<16$ the detection image is saturated, resulting in an incorrect aperture correction even though the $K_s$ band photometry is unsaturated.  This is manifested as an apparent systematic trend with magnitude between the COSMOS catalog and 2MASS photometry.  However, no trend is visible for objects fainter than $i^+>16$ and with measurements made directly on the $K_s$ band image.

\begin{figure}
\begin{center}
\includegraphics[scale=0.35]{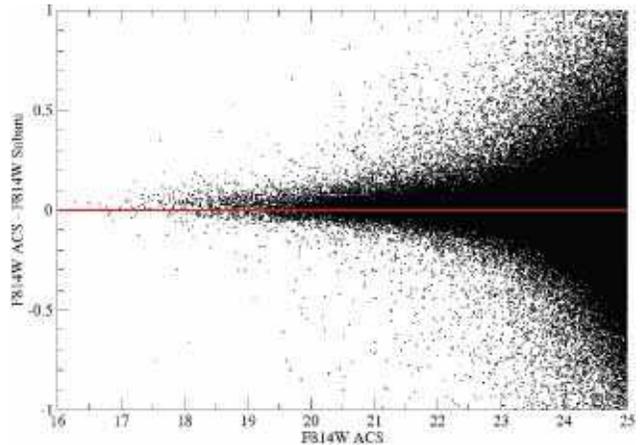}
\caption{The offset between the ACS F814W band photometry and the Subaru photometry is shown as a function of magnitude after applying the offsets in Table \ref{t:zp-corr}.  The Subaru $i^+$ and $z^+$ bands were used to estimate F814W photometry.  The scatter and zero point error is within the expected error of the color conversion. \label{f:F814W-offset}}
\end{center}
\end{figure}

\begin{deluxetable}{lcc}
\tabletypesize{\scriptsize}
\tablecaption{RMS Variation in the Photometric Zero Point as a Function of Position Between Surveys\label{t:zpvar}}
\tablehead{
\colhead{COSMOS Filter} & \colhead{CFHT-LS} & \colhead{SDSS}
}
\startdata
$u^*$	&	0.014	&	\nodata\tablenotemark{1}	\\
$B_J$	&	0.009	&	0.006	\\
$g^+$	&	0.010	&	0.013	\\
$V_J$	&	0.010	&	0.020	\\
$r^+$	&	0.013	&	0.009	\\
$i^+$	&	0.007	&	0.012	\\
$z^+$	&	0.010	&	0.017	\\
\enddata
\tablenotetext{1}{A comparison between SDSS $u$ and CFHT $u^*$ is not possible because there is no simple linear relationship between these filters.}
\end{deluxetable}

	The ACS photometry has the best relative calibration of all the data sets due to the lack of atmospheric absorption. So the excellent agreement between the ACS and Subaru $i^+$ photometry as a function of position indicates that the variations seen between COSMOS and the CFHT-LS and SDSS are largely due to flat fielding or sky subtraction errors in those other surveys.  Nevertheless, our photometry appears to be constant across the COSMOS field to better than 1\%, meeting the science requirement for large scale structure studies. 
	
\subsection{Galactic Extinction Correction}

\begin{deluxetable}{lcc}
\tabletypesize{\scriptsize}
\tablecaption{Galactic Extinction Corrections\label{t:galext}}
\tablehead{
\colhead{Filter} & \colhead{k$_\lambda$\tablenotemark{*}} & \colhead{Median A$_\lambda$} 
}
\startdata
$u$ & 4.996724 & 0.097436\\
$u^*$ & 4.690237 & 0.091460\\
$B_j$ & 4.038605 & 0.078753\\
$g$ & 3.791856 & 0.073941\\
$g^+$ & 3.738239 & 0.072896\\
$V_j$ & 3.147140 & 0.061369\\
$r$ & 2.649158 & 0.051659\\
$r^+$& 2.586050 & 0.050428\\
$i$ & 1.989881 & 0.038803\\
$i^+$ & 1.922693 & 0.037493\\
$i^*$ & 1.922912 & 0.037497\\
F814W & 1.803909 & 0.035176\\
NB816 & 1.744951 & 0.034027\\
$z$ & 1.467711 & 0.028620\\
$z^+$ & 1.435914 & 0.028000\\
$K_s$ & 0.340677 & 0.006643\\
\enddata
\tablenotetext{*}{A$_\lambda$ = k$_\lambda$*e(B-V)}
\end{deluxetable}

	The median galactic extinction in the COSMOS field is e(B-V)$=0.0195\pm0.006$, which corresponds to an extinction of $0.10\pm0.03$ magnitudes in the $u^*$ band.  The estimated galactic extinction from \citet{1998ApJ...500..525S} is provided for each object in the COSMOS catalog.  A photometric correction for each band can be determined from the galactic extinction multiplied by a filter dependent factor.  These factors are given in Table \ref{t:galext} for filters used on the COSMOS field.  These band-pass dependent factors are calculated by integrating the filter response function against the galactic extinction curve provided by \cite{2000A&A...363..476B}, originally taken from \citet{allen-astrophysquant}.
	
\subsection{Absolute Photometric Zero-Point Corrections \label{s:zp-corr}}
\begin{deluxetable}{lcc}
\tabletypesize{\scriptsize}
\tablecaption{Photometric Offsets Measured from Other Surveys\label{t:zp-corr-CFHTLS}}
\tablehead{
\colhead{Filter} & \colhead{Offset to CFHT-LS} &  \colhead{Offset to SDSS}
}
\startdata
$u^*$	&	0.035	&	\nodata \tablenotemark{*} \\
$B_J$	&	0.125	&	0.11\\
$g^+$	&	-0.096	&	-0.12\\
$V_J$	&	0.040	&	0.03\\
$r^+$	&	-0.080 	&	-0.07\\
$i^+$	&	-0.093	&	-0.10\\
$z^+$	&	-0.032	&	-0.03\\
\enddata
\tablenotetext{*}{There is no linear relationship between SDSS $u$ and CFHT $u^*$}
\end{deluxetable}

	With typical overheads of 15 minutes per standard it is extremely difficult to obtain a sufficient number of standard stars on Suprime-Cam.  With three to five standards per band \citep{tanaguchi-cosmos}, our standard star calibrations are accurate to $\pm 0.05$ magnitudes (see \citet{aussel-photom}).  These offsets are larger than desired for accurate photometric redshifts. 

	Comparisons with the CFHT-LS and SDSS yield the zero point offsets given in Table \ref{t:zp-corr-CFHTLS}.  These were estimated by comparing total magnitudes for point sources between $21^{st}$ and $24^{th}$ magnitude for CFHT-LS and 18th and 21st magnitude for SDSS converted to the COSMOS-AB photometric system.  The color conversions between the surveys are given in Table \ref{t:color-conv-SDSS} \& \ref{t:color-conv-CFHT} and discussed further in \citet{aussel-photom}.
	
	The zero point corrections from the two surveys are consistent with one another and have an RMS amplitude of $\pm 0.06$ magnitudes, which is slightly larger than the expected error.  However, the zero points calculated in this way disagree with the ACS F814W photometry by -0.118 magnitudes and fail to produce photometric redshifts free from systematic errors.  Furthermore, \citet{2006astro.ph..3217I} find that the CFHT-LS zero-points are inaccurate at the 0.05 magnitude level.

 	Considering the number of present and ongoing observations of COSMOS, obtaining spectrophotometric standards in the field is a reasonable long term solution to obtaining better quality photometry \citep{2001A&A...365..681W}.  In the interim we are forced to rely on the existing calibrations and spectra of galaxies to re-calibrate the photometric zero points for photometric redshifts.

\begin{deluxetable}{lc}
\tabletypesize{\scriptsize}
\tablecaption{Photometric Offsets Calculated with Spectroscopic Redshfits\label{t:zp-corr}}
\tablehead{
\colhead{Filter} & \colhead{Offset}
}
\startdata
$u$ &	0.0		\\
$u^*$	&	-0.084	\\
$B_J$	&	0.189	\\
$g$ &	0.01		\\
$g^+$	&	-0.090	\\
$V_J$	&	0.04		\\
$r$ &	-0.033	\\
$r^+$	&	-0.040	\\
$i$ &	-0.037	\\
$i^+$	&	-0.020	\\
$i^*$		&	-0.005	\\
NB816	&	-0.072	\\
F814W	&	0.000	\\
$z$ &	-0.037	\\
$z^+$	&	0.054	\\
$K_s$&	-0.097\tablenotemark{*}  \\
\enddata
\tablenotetext{*}{Measured from 2MASS, not from spectroscopic redshifts}
\end{deluxetable}

\begin{figure*}
\begin{center}
\begin{tabular}{cc}
\includegraphics[scale=0.3]{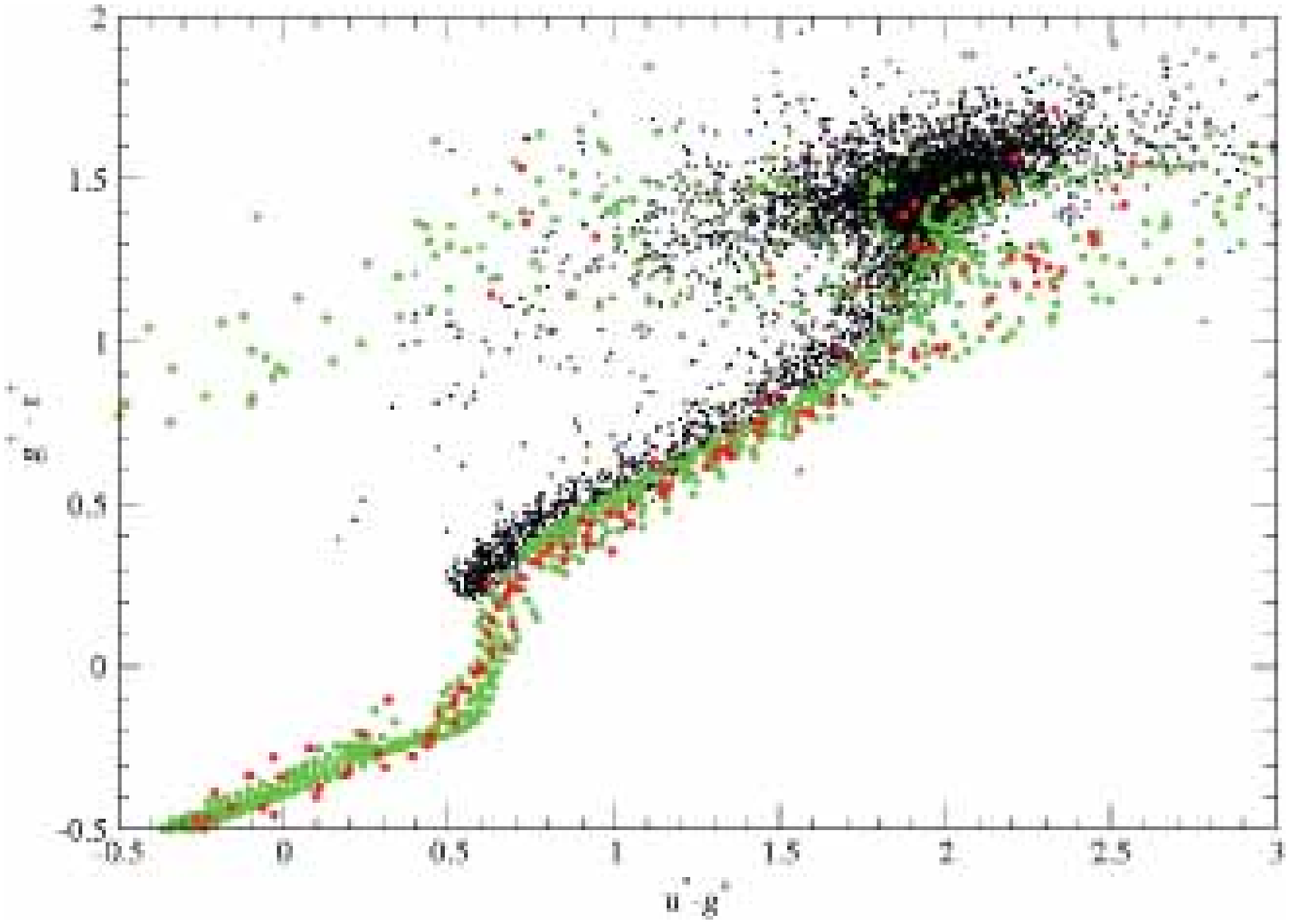}&\includegraphics[scale=0.3]{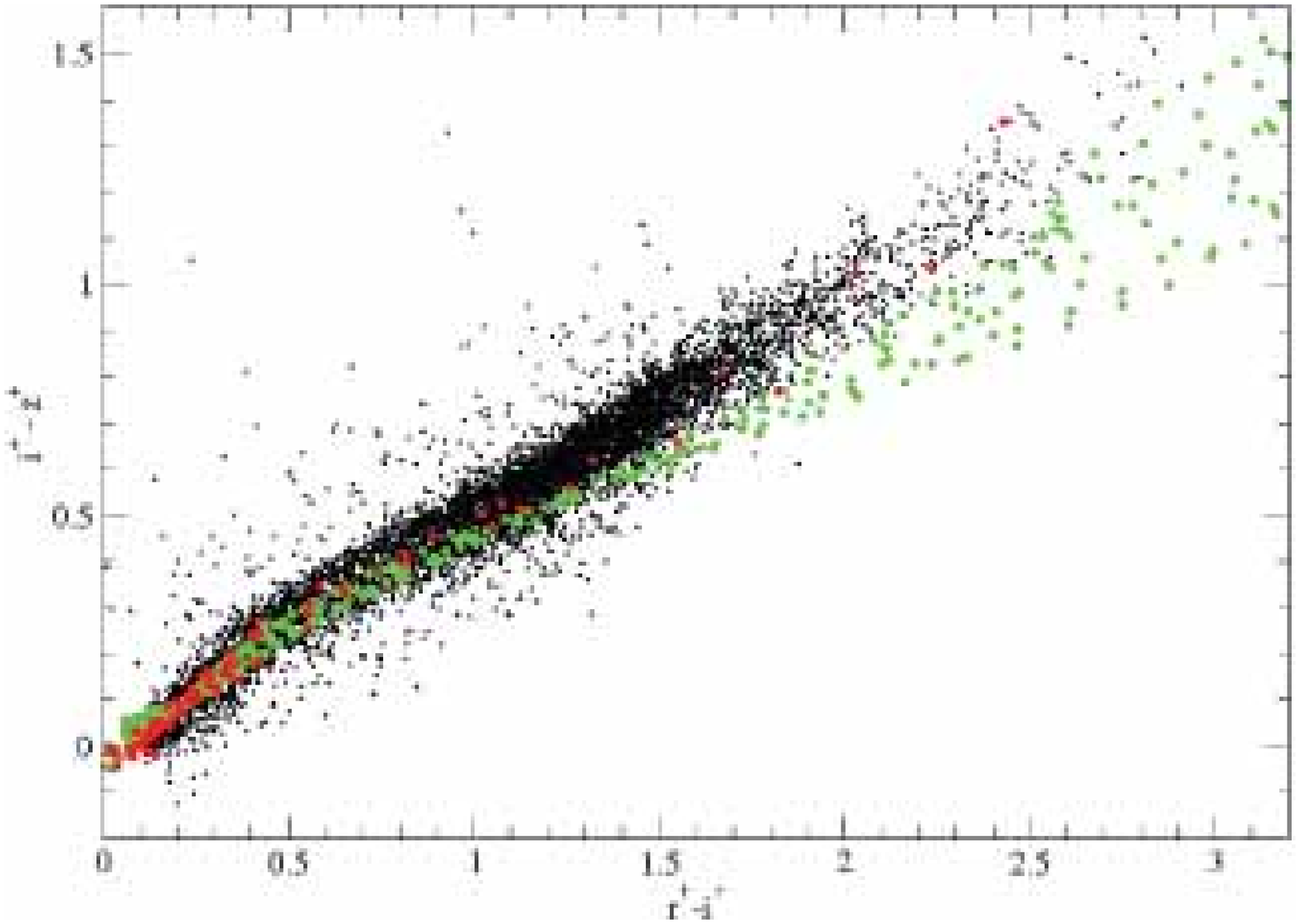}\\
\includegraphics[scale=0.3]{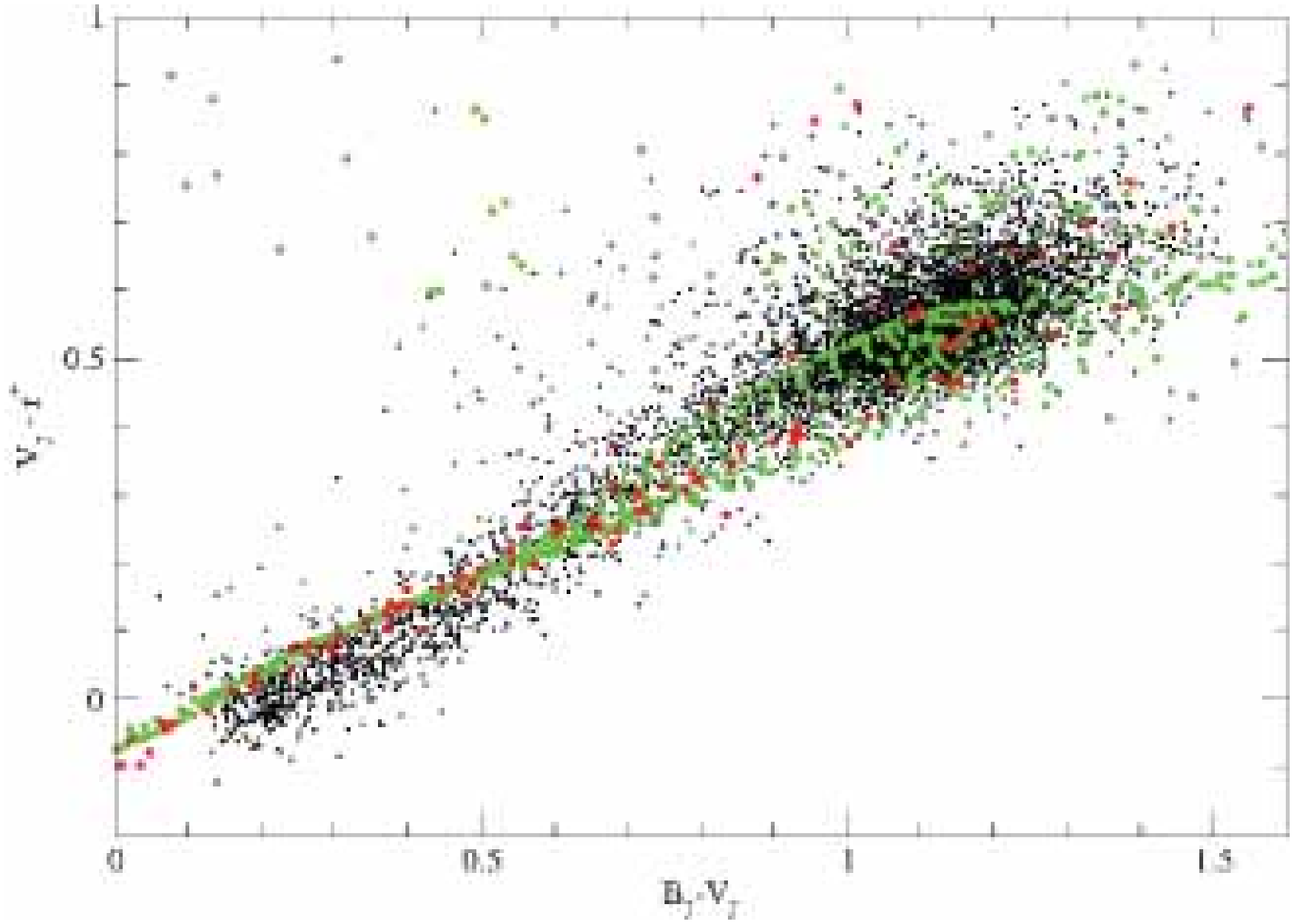}&\includegraphics[scale=0.3]{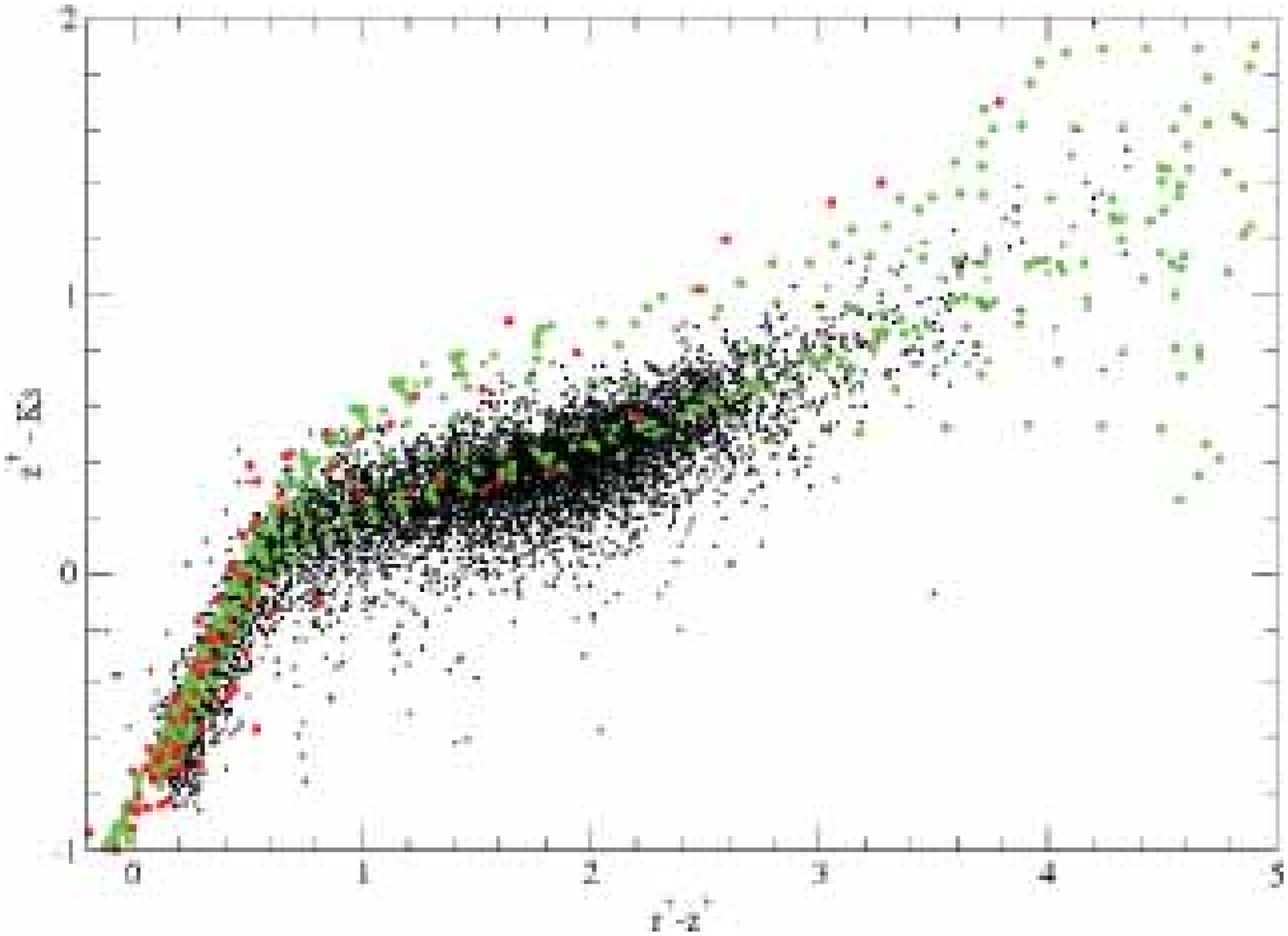}\\
\end{tabular}
\caption{Four Color-Color plots are shown for point sources in the COSMOS field (black) along with the expected colors from the BaSel 3.1 \citep{2002A&A...381..524W} and NextGen \citep{2005MSAIS...7..140H, 2005tdug.conf..565B, 2002ASPC..274...95H} stellar libraries at the expected median metallicity of [Fe/H]=-0.4 (See \citet{rich-stars}) (green), and the \citet{pickles} spectral library ([Fe/H]$\simeq0$) (red).  Notice that the Pickles library predicts systematically redder colors than the observations in the UV due to the high metallicity of these stars. \label{f:star-color}}
\end{center}
\end{figure*}

	These zero point offsets are calculated by fitting the measured photometry, corrected for galactic extinction, to galaxy templates at the known redshift and calculating the offset between the measured and expected photometry.  To avoid  systematic variations due to calibration errors in the galaxy templates, offsets between bands are calculated in rest frame wavelength bins of 100\AA\  separately for each template, then combined in a weighted average.  This is effective because calibration errors in the template will create offsets which vary as a function of rest wavelength in the same way for all bands, while zero point offsets between bands will be constant as a function of wavelength.  Since these offsets are relative, the F814W is used as the reference for the absolute zero point.  The offsets calculated using this method are given in Table \ref{t:zp-corr}.  	

	This method is effective with a large sample of spectroscopic redshifts covering a large range in redshift along with photometry taken in many adjacent bands.  For COSMOS there are insufficient data to correct the $K_s$ band due to the lack of photometric data between $0.9\mu$m and $2.2\mu$m.  As a result we rely on the 2MASS zero point for those data.  Although the individual exposures were tied to 2MASS, an offset of -0.097 magnitudes is measured between the catalog photometry and the 2MASS catalog.  This offset is likely an aperture correction between the photometry measured on the individual CTIO and KPNO exposures and the 2MASS catalog. 
	
	With the exception of the Suprime-Cam $B_J$ band, which has known calibration problems (see \citet{aussel-photom}), the offsets are within the expected error of $\pm 0.05$ magnitudes.  After applying these offsets no systematic trend is measurable between photometric and spectroscopic redshifts for 842 objects between $0<z<1.2$.   In addition, the $i^+$ and $z^+$ band agree with the F814W photometry to 0.007 magnitudes and the offsets between the COSMOS and CFHT-LS photometry agree with those in \citet{2006astro.ph..3217I} after applying color corrections.  Furthermore, after applying the offsets the colors of stars agree with the predicted colors extremely well (see Section \ref{s:stars}).  These tests indicate the zero points are within 0.01 magnitude of the true AB zero points after applying these offsets.
	
		These corrections were not applied to the released catalog since they can not be verified with external calibration sources at this time.  However, we recommend applying the photometric redshift offsets for the best possible photometry and colors.

\subsection{Star Colors}\label{s:stars}

	Stars are a good diagnostic of color accuracy because they form a tight sequence in most optical and near-IR color-color plots.  Offsets as small as a few hundredths of a magnitude are visible when comparing expected and measured star colors.  Furthermore, star colors are sensitive to the filter transmission profiles, providing a valuable check of the instrumental performance \citep{aussel-photom}.  
	
	Even at the resolution of HST, compact galaxies and quasars contaminate a star selection based on morphology.  These objects create scatter in color-color plots, obfuscating the stellar locus.  The BzK color-color diagram provides a much cleaner star selection \citep{2004ApJ...617..746D}, and was used to select the objects plotted in Figure \ref{f:star-color}.  The BzK method is biased against faint blue stars due to the shallow $K_s$ band data, however, the effect is minimal since only objects with greater than 10$\sigma$ detections are plotted.  

	Stellar libraries e.g. \citep{pickles} typically contain solar metallicity stars ([Fe/H]$\simeq$0), which differ in color from the dominant sub-solar metallicity ([Fe/H]$\simeq$-0.4), thick disk population in the COSMOS field (see \citet{rich-stars}).  This results in small offsets, especially in the ultraviolet where metal line absorption will cause higher metallicity stars to appear redder.  
	
	Figure \ref{f:star-color} shows four different color-color plots for stars along with colors for the \citet{pickles} library and a combination of the BaSel 3.1 \citep{2002A&A...381..524W} and NextGen theoretical libraries \citep{2005MSAIS...7..140H, 2005tdug.conf..565B, 2002ASPC..274...95H} at [Fe/H]=-0.4.  The star colors agree extremely well in all color-color plots with an offset between the median expected and actual colors of less than 0.02 magnitudes in most bands.  A correction of -0.05 magnitudes is indicated by the stellar track for $u^*$, and systematic differences between the predicted and actual colors for $B_J$ and $z^+$ are observed.  These are likely due to our limited measurements of the filter throughput \citep{aussel-photom}.
	
\subsection{Number Counts}

	Differential number counts of galaxies are a simple but powerful measurement of the geometry of space, the evolution of the galaxy population and the evolution of structure in the universe.   They also provide a valuable check on data quality because they are sensitive to photometric calibration errors, detection completeness, and spurious detections.  Figure \ref{f:Icounts} shows the I band number counts from the COSMOS catalog compared to other surveys.  The numbers are in good agreement, however the COSMOS $i^+$ band counts are higher than the F814W ACS counts at magnitudes fainter than 24.  This is likely due to the lack of sensitivity at low surface brightness in ACS compared to Subaru and some contamination from spurious sources in the ground based catalog.  The $i^+$ band galaxy number counts are given in Table \ref{t:iband-counts}; stars have been removed using a cut in ground based FWHM since ACS does not cover the entire 2 square degrees.  

\begin{figure}
\begin{center}
\includegraphics[scale=0.3]{f12.eps}
\caption{I band number counts are shown for the COSMOS ground based catalog, the COSMOS F814W weak lensing catalog \citet{acs-weak-lensing}, the Hawaii-Hubble deep field (H-HDF-N) \citep{capak-HDF}, the Hubble Deep Field North (HDF-N) \citep{1996AJ....112.1335W,2001MNRAS.323..795M}, the Hubble Deep Field South (HDF-S) \citep{2001MNRAS.323..795M},  the Herschel Deep Field \citep{2001MNRAS.323..795M}, the SDSS \citep{2001AJ....122.1104Y}, the Canada France Deep Fields (CFDF) \citet{2003A&A...410...17M}, and the Canada-France-Hawaii Telescope Legacy Survey (CFHT-LS) \citep{mccracken_clustering}.  Our counts are in good agreement with other surveys up to our 80\% completeness limit at $i^+ = 26.5$. \label{f:Icounts}}
\end{center}
\end{figure}

\begin{figure}
\begin{center}
\includegraphics[scale=0.3]{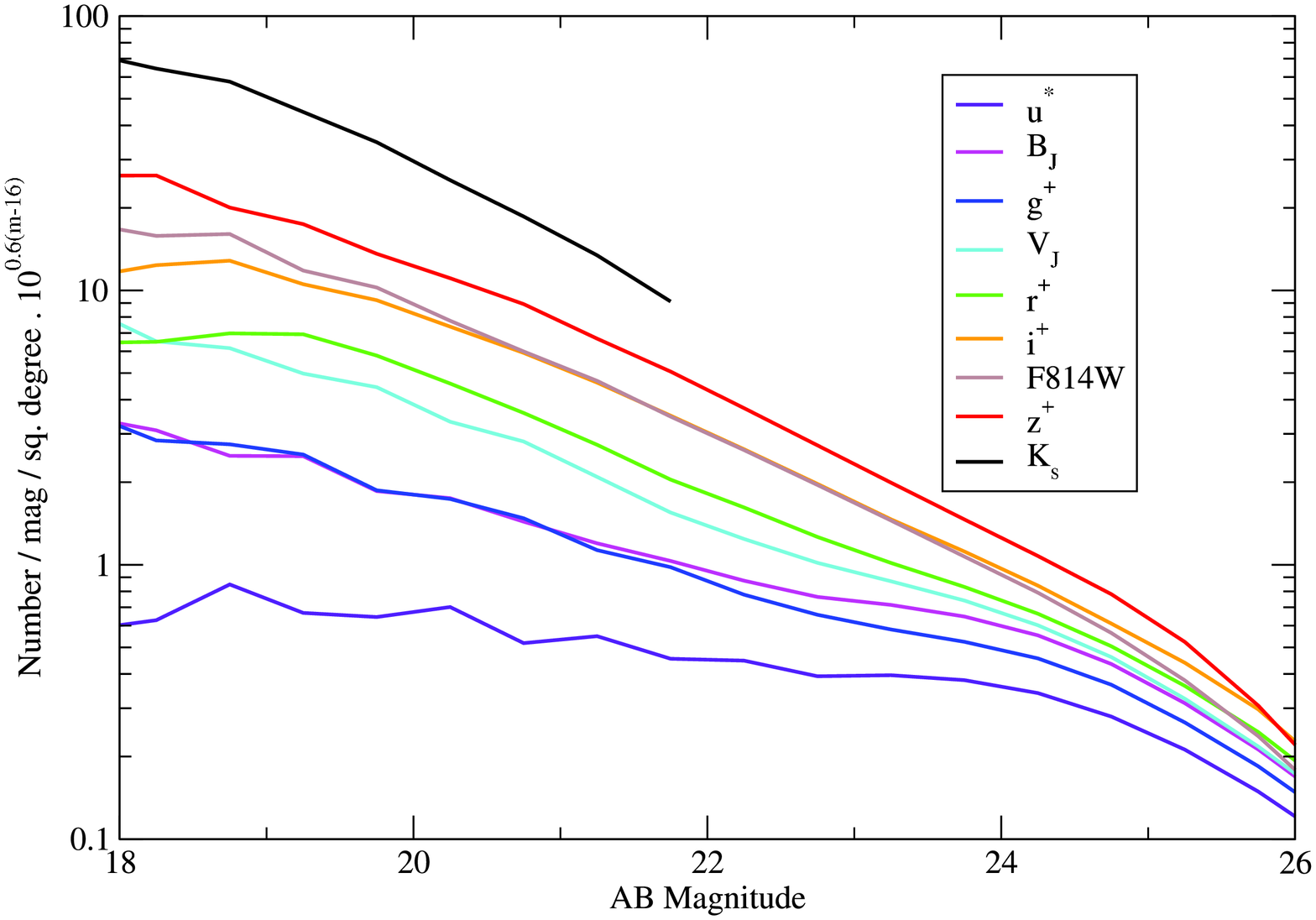}
\caption{Number counts from the $i^+$ detected catalog are shown for $u^*$, $B_J$, $g^+$, $V_J$, $r^+$, $i^+$, F814W, $z^+$, and $K_s$ bands divided by a Euclidian slope.  The $u^*$ counts are nearly flat while $K_s$ band counts are very steep.  A combination of a $\Omega_m=0.3$, $\Omega_v=0.7$ geometry, galaxy evolution, and redshifting is responsible for the band dependent behavior. \label{f:numcounts}}
\end{center}
\end{figure}

\begin{deluxetable}{ccc}
\tabletypesize{\scriptsize}
\tablecaption{$i^+$ Band Number Counts\label{t:iband-counts}}
\tablehead{
\colhead{$i^+$ AB} & \colhead{N deg$^{-2}$ mag$^{-1}$ } & \colhead{Poisson Error}}
\startdata
15.625 & 6.534683 & 3.772801 \\ 
15.875 & 4.350995 & 3.076618 \\ 
16.125 & 4.345201 & 3.072521 \\ 
16.375 & 4.339069 & 3.068185 \\ 
16.625 & 12.997791 & 5.306326 \\ 
16.875 & 38.932047 & 9.176372 \\ 
17.125 & 49.664184 & 10.355698 \\ 
17.375 & 75.444603 & 12.752466 \\ 
17.625 & 114.0366 & 15.664132 \\ 
17.875 & 154.62239 & 18.222423 \\ 
18.125 & 265.76389 & 23.866302 \\ 
18.375 & 359.31933 & 27.722088 \\ 
18.625 & 525.00773 & 33.473271 \\ 
18.875 & 706.95209 & 38.799037 \\ 
19.125 & 871.08483 & 43.019813 \\ 
19.375 & 1121.62 & 48.812246 \\ 
19.625 & 1538.1477 & 57.204325 \\ 
19.875 & 1978.0351 & 64.862326 \\ 
20.125 & 2487.0499 & 72.771793 \\ 
20.375 & 3150.2535 & 81.970061 \\ 
20.625 & 4084.8722 & 93.296944 \\ 
20.875 & 4900.1737 & 102.1979 \\ 
21.125 & 6309.9719 & 116.11697 \\ 
21.375 & 7656.9391 & 127.9895 \\ 
21.625 & 9710.6197 & 144.03897 \\ 
21.875 & 12951.105 & 166.68489 \\ 
22.125 & 15769.274 & 184.26274 \\ 
22.375 & 18613.365 & 199.97038 \\ 
22.625 & 22660.648 & 220.97698 \\ 
22.875 & 27245.578 & 242.54972 \\ 
23.125 & 32943.681 & 267.97654 \\ 
23.375 & 40428.641 & 297.86588 \\ 
23.625 & 49326.475 & 328.82855 \\ 
23.875 & 60011.205 & 363.35744 \\ 
24.125 & 74117.459 & 405.50413 \\ 
24.375 & 90498.908 & 450.39941 \\ 
24.625 & 109643.14 & 498.12066 \\ 
24.875 & 132088.39 & 548.23564 \\ 
25.125 & 161449.58 & 611.05628 \\ 
25.375 & 194924.55 & 678.02406 \\ 
25.625 & 232494.58 & 750.00984 \\ 
25.875 & 264905.22 & 808.54775 \\ 
26.125 & 290603.29 & 855.89733 \\ 
26.375 & 313097.56 & 910.75629 \\ 
\enddata
\end{deluxetable}
	
	Figure \ref{f:numcounts} show the number counts for the COSMOS bands normalized to a slope of 0.6, which is expected in a static Euclidian universe.  In a universe with no galaxy evolution and a cosmology of $\Omega_{m}=0.3$, $\Omega_{v}=0.7$, the number count slope should be below 0.6 in all bands.  Furthermore, galaxies have a flat to red spectrum in units of F$_\nu$, so the k-correction is positive in AB magnitudes.  This results in galaxies fading faster with redshift than they would from simple luminosity distance, further flattening the number count slope.
	
	  The slope of the normalized number counts in Figure \ref{f:numcounts} steepen with increasing wavelength indicating both the universe and the galaxy population is evolving.  The UV luminosity of a typical galaxy increases with redshift due to increased star formation activity \citep{1999AJ....118..603C,1996ApJ...460L...1L,2005ApJ...631..126D}.   This galaxy evolution counters the effects of an evolving universe, resulting in a $u^*$ band slope close to 0.6.  In the near-IR, there is little galaxy evolution \citep{2005ApJ...631..126D} so the counts in the $K_s$ band are far below a slope of 0.6.

\section{Conclusions}
Deep ground based data were presented for the COSMOS field in 15 bands between $0.3\mu$m-$2.2\mu$m along with an $i$ band selected catalog of objects.  We show that these data have an extremely high level of photometric consistency necessary for scientific pursuits such as large scale structure studies.  The expected zero-point variations are $<1$\% across the field; this level of photometry was achieved by constructing flat fields directly from object fluxes rather than using sky or dome flats.

Variations in the point spread function (PSF) are a significant source of uncertainty in ground based photometry.  This is due to the fact that the non-Gaussian portion of the PSF is difficult to match across multiple bands, resulting in 2-5\% errors in color measurement if uncorrected. The variation was minimized by adjusting the Gaussian filter used for PSF matching to provide the same fraction of light to fall in a 3\asec aperture in all bands.  However, this means that only the 3\asec aperture photometry is reliable for colors.  Photometry measured in other aperture sizes should be corrected for variations in the PSF.

An $i^+$ band selection was used for the multi-band catalog after analyzing the tradeoffs between a $\chi^2$ and $i$ band detected catalog.  The decrease in resolution and increased problems with a $\chi^2$ catalog outweigh the benefits of increased sensitivity and pan-chromatic completeness.  The resulting catalog is 91\% complete at $i^+=25.0$, 87\% complete at $i^+=26.0$ and 50\% complete at $i^+=27.4$. 	

Our photometric zero-points measured from standard stars \citep{aussel-photom} are accurate to 5\%.  These zero-point offsets can be significantly improved by using galaxies with known redshifts to adjust the zero-point calibration.  The corrected zero-points appear accurate to better than 2\% based on star colors; this will be verified with future internal spectrophotometric standards.  

Number counts measured from the $i^+$ band selected catalog agree well with previous studies.  The effects of galaxy evolution and an expanding universe are clearly visible in the pan-chromatic counts, demonstrating the importance of multi-color surveys.

\acknowledgements
We would like to thank the COSMOS team \url{http://cosmos.astro.caltech.edu/}, the staff at Caltech, CFHT, CTIO, KPNO, NAOJ, STSCI, Terapix, and the University of Hawaii for supporting this work and making it possible.  Support for this work was provided by NASA grant HST-GO-09822 and NSF grant OISE-0456439.

FLAMINGOS was designed and constructed by the IR instrumentation group (PI: R. Elston) at the University of Florida, Department of Astronomy, with support from NSF grant AST97-31180 and Kitt Peak National Observatory"

Funding for the Sloan Digital Sky Survey (SDSS) has been provided by the Alfred P. Sloan Foundation, the Participating Institutions, the National Aeronautics and Space Administration, the National Science Foundation, the U.S. Department of Energy, the Japanese Monbukagakusho, and the Max Planck Society. The SDSS Web site is http://www.sdss.org/.

The SDSS is managed by the Astrophysical Research Consortium (ARC) for the Participating Institutions. The Participating Institutions are The University of Chicago, Fermilab, the Institute for Advanced Study, the Japan Participation Group, The Johns Hopkins University, Los Alamos National Laboratory, the Max-Planck-Institute for Astronomy (MPIA), the Max-Planck-Institute for Astrophysics (MPA), New Mexico State University, University of Pittsburgh, Princeton University, the United States Naval Observatory, and the University of Washington.

D. Thompson acknowledges support from NASA through a Long Term Space Astrophysics grant (NAG5-10955, NRA-00-01-LTSA-064).

This work is based (in part) on data products produced at TERAPIX data center located at the Institut d'Astrophysique de Paris.

\end{document}